\documentclass[aps,prd,twocolumn,eqsecnum,amssymb,amsmath,showpacs,a4paper, superscriptaddress]{revtex4-1}

\usepackage{graphicx}
\usepackage{amsfonts}
\usepackage{amsmath}
\usepackage{hyperref}
\usepackage{units}
\usepackage[latin1]{inputenc}    		
\usepackage{color}
\usepackage{dcolumn}
\usepackage{bm}
\usepackage{float}
\usepackage{cleveref}
\usepackage[normalem]{ulem}
\usepackage{appendix}
\usepackage{mathtools}

\graphicspath{ {images/} }
\usepackage{mathrsfs}
\usepackage{amssymb}
\usepackage{dsfont}
\usepackage{enumitem}
\usepackage{gensymb}
\usepackage{bm}

\crefname{section}{§}{§§}
\Crefname{section}{§}{§§}
\usepackage{mathtools}
\usepackage{chngcntr}
\counterwithout{equation}{section}

\newcommand{\cmmnt}[1]{\ignorespaces}
\renewcommand{\vec}[1]{\mathbf{#1}}
\newcommand{\be} {\begin{equation}}
\newcommand{\ee} {\end{equation}}
\newcommand{\bsub}{\begin{subequations}}
\newcommand{\esub}{\end{subequations}}
\newcommand{\bea}{\begin{eqnarray}}
\newcommand{\eea}{\end{eqnarray}}

\usepackage[normalem]{ulem}

\def\be{\begin{equation}}
\def\ee{\end{equation}}
\def\ba{\begin{align}}
\def\ea{\end{align}}

\newcommand{\silke}[1]{{\textcolor{red}{{#1}}}}

\newcommand{\p}{p}
\newcommand{\IN}{\mathrm{L}}
\newcommand{\OU}{\mathrm{R}}
\newcommand{\tp}{\mathit{tp}}

\newcommand{\SH}{\mathcal{F}}

\begin{document}

\title{Superradiance in dispersive black hole analogues}

\author{Sam Patrick}
 \email{sampatrick31@googlemail.com}
 \affiliation{%
 Mathematical Sciences, Durham University\\
 Durham, DH1 3LE, United Kingdom
}
\author{Silke Weinfurtner}%
 \email{silkiest@gmail.com}
\affiliation{%
 School of Mathematical Sciences, University of Nottingham\\
 Nottingham, NG7 2FD, United Kingdom
}%
\affiliation{
Centre for the Mathematics and Theoretical Physics of Quantum Non-Equilibrium Systems, \\
University of Nottingham, Nottingham, NG7 2FD, United Kingdom
}

\date{\today}

\begin{abstract}
Wave equations containing spatial derivatives which are higher than second order arise naturally in the context of condensed matter systems. The solutions of such equations contain more than two modes and consequently, the range of possible interactions between the different modes is significantly enhanced compared to the two mode case. We develop a framework for analysing the different mode interactions based on the classical turning points of the dispersion relation. We then apply this framework to the scattering of deep water gravity waves with a draining bathtub vortex, a system which constitutes the analogue of a rotating black hole in the non-dispersive limit.
In particular, we show that the different scattering outcomes are controlled by the light-ring frequencies, a concept routinely applied in black hole physics, and two new frequencies which are related to the strength of dispersion.
We find that the frequency range in which the reflected wave is superradiantly amplified appears as a simple modification to the non-dispersive case.
However, the condition to observe this amplification is complicated by the fact that a superradiant mode can be reflected back into the system by scattering with one of the additional modes.
We provide estimates for the reflection coefficients in the full dispersive regime.
\end{abstract}

\maketitle

\section{Motivation}

Analogue models of gravity are an exciting area of research that have attracted much attention on both the theoretical and experiment fronts in the past two decades.
The analogue gravity programme aims to shed light on physical processes arising in gravitational physics and condensed matter systems alike
\cite{unruh1981experimental,visser1993lorentzian,barcelo2011analogue},
and as such has become an active area of inter-disciplinary research.
A particularly promising analogue gravity system consists of water waves propagating on open channel flows. Small perturbations with wavelengths larger than the water depths, which are commonly referred to as shallow water waves, exhibit dynamics which can be mapped onto wave propagation on a curved spacetime, e.g. in the vicinity of non-rotating~\cite{rousseaux2008observation,weinfurtner2011measurement,weinfurtner2013classical,euve2016observation,euve2020scattering} and rotating black holes~\cite{torres2018application}. The effective spacetime geometry is fully determined by fluid parameters~\cite{schutzhold2002gravity}, and thus by setting up specific fluid flows, one can mimic a variety of analogue spacetime geometries. The overarching goal within this field of research is to study effects predicted within classical and quantum field theory on curved spacetime geometries in analogue gravity systems. 

For example, amongst the first successful experimental realisations of an analogue gravity system were those concerning gravity waves on an effectively one-dimensional open channel flow, with varying depth, exhibiting effective black and white hole horizons. These experimental explorations lead to the detection of the classical equivalent of Hawking radiation, see for example \cite{rousseaux2008observation,weinfurtner2011measurement,weinfurtner2013classical,euve2016observation,euve2020scattering}. Consequently, there exists a large body of work in the literature devoted to understanding the Hawking effect in surface wave analogues, adapting the naive analogy to experimental reality, e.g.~taking dispersive effects into account, \cite{corley1996spectrum,macher2009black,finazzi2012hawking,coutant2012black,coutant2014hawking,robertson2016scattering,coutant2014undulations,coutant2016imprint}. 

The focus here is on analogue gravity systems exhibiting two spatial dimensions, allowing the study of processes associated with rotating black holes in a controlled laboratory setting.
In particular modelling efforts are concentrating on rotating, draining vortex flows. These models possess both a horizon and an ergosphere and thus have the potential to mimic a variety of rotating black hole processes,  \cite{dolan2011AB,dolan2012resonances,dolan2013scattering}. There are two such processes that have recently been demonstrated in a laboratory setup -- ringdown and superradiance -- which we introduce next.

Perturbed black holes relax toward equilibrium via the emission of quasinormal modes (QNMs), solutions of the equation of motion with complex frequencies which obey dissipative boundary conditions.
QNM emission (or ringdown) is also expected to occur around draining vortices \cite{berti2004qnm,cardoso2004qnm} and the real part of the QNM spectrum was recently measured in an experiment \cite{torres2018application}.
It has since been argued that the QNM spectrum could be used to extract information about the fluid's velocity using a flow characterisation method based on black hole spectroscopy \cite{torres2019analogue}.
This demonstrates how techniques borrowed from black hole physics can be used to develop new insights and methods to study fluid systems, and thus testifies to the two-way utility of analogue gravity studies.

Rotational superradiance, the subject of our study, is an effect that is expected to occure in the vicinity of rotating black holes and vortex flows alike \cite{basak2003superresonance,basak2003reflection,richartz2015rotating}.
Superradiance is an energy enhancement effect, in which an incident wave is amplified during a scattering event, thereby extracting energy from the system (see \cite{bekenstein1998many,brito2015superradiance} for a review).
With origins in quantum mechanics
\cite{ginzburg1947doppler,ginzburg1993v,dicke1954coherence,zeldovich1971generation,zeldovich1972amplification},
superradiance appears under different guises in many disciplines. For example, it is related to over-reflection in fluid mechanics
\cite{mckenzie1972reflection,acheson1976overreflexion,kelley2007inertial,fridman2008overreflection}.
The name superradiance, however, has become most associated with superradiance around rotating black holes,
\cite{penrose1971extraction,misner1972stability,starobinsky1974waves,starobinsky1974electro}
where it played a key role in the early development
of black hole thermodynamics~\cite{hawking1974explosions,bekenstein1994entropy}.
More recently, proposals have been outlined to search for physics beyond the standard model using black hole superradiance \cite{brito2017gravitational,baumann2019probing,siemonsen2020gravitational}.
The first direct detection of rotational superradiance was performed in \cite{torres2017rotational} using a draining vortex flow and, although the analogy to black hole physics using surface waves is mathematically precise only in the shallow water regime, the amplification was in fact measured closer to the deep water regime where the system is strongly dispersive. The strong dispersive regime requires one to work with the full dispersion relation, instead of the weak dispersive regime, where only next order correction to the non-dispersive regime are being taken into account.

However, little to no work has been carried out on the theoretical modelling of superradiant scattering processes in regimes readily accessible for hydrodynamic rotating black hole experiments. With experiments on-going, there is a need for improvement in the theoretical modelling. 
Below we present an analytical study of scattering of surface waves from a hydrodynamic rotating black hole in the full dispersive regime.

\section{Methodology}

We apply a combination of multiple scale analysis and matched asymptotics techniques, to obtain an (approximate) analytic and globally defined solution for scattering processes arising within analogue rotating and non-rotating black holes in two-spatial dimensions. 

We first apply the Wentzel-Kramers-Brillouin (WKB) method, which is a particular case of what is more generally referred to as multiple scale analysis \cite{buhler2014waves}.
The basic idea behind the WKB approximation is to assume that the solutions can be split into independent variables; a slowly varying amplitude and rapidly varying phase. As we demonstrate below, this ansatz allows us to recast the complicated problem of scattering processes of dispersive waves in terms of a much simpler problem of scattering of point-like particles. 
This method is precisely analogous to semiclassical approximations in quantum mechanics \cite{berry1972semiclassical}, and is also known as ray-tracing (e.g. in plasma physics \cite{tracy2014ray}), where the light rays are in direct correspondence with the trajectories of semi-classical particles.
The two separate notions of wave-like and particle-like behaviour become equivalent in the limit of small wavelengths (or in our case, large azimuthal numbers) and thus, the methods laid out in this work are expected to yield increasing accuracy for the modes in the system with high angular momentum. Indeed, similar methods based on a WKB approximation have already been shown to accurately predict experimental observations e.g. the black hole ringdown behaviour from hydrodynamic rotating black holes~\cite{torres2018application}. 

However, as commonly known, the approximate WKB solutions become singular at turning points, which are the locations where a classical particle changes direction.
In optics, these locations are known as caustics \cite{buhler2014waves}.
This failure is to be expected in a sense, since the WKB solution can only account for adiabatic changes in each mode. Thus if one wants to study non-adiabatic process, in particular wave scattering between modes, the WKB method is expected break down or nothing interesting would happen.
The presence of turning points indicates the presence of non-trivial mode interaction.
When the system has no real turning points, one usually has to look to more intricate methods beyond WKB, e.g. in the presence of complex turning points \cite{coutant2016imprint}.

The focus in this work is to investigate the different possible outcomes of wave scattering in dispersive systems by studying modes propagation around turning points.
To this end, we first construct WKB solutions which are valid sufficiently far away from the turning points then solve the wave equation exactly around the turning points.
In order to construct globally defined solutions, we apply the method of matched or intermediate asymptotics to patch together the WKB solutions.

Within our approach, we are circumventing one of the major difficulties in applying standard black hole techniques to dispersive analogue systems. Analogue horizons are well-defined in the non-dispersive regime and occur at locations when the perturbation speed equals the speed of the fluid flow. However, in dispersive systems the horizon is not a well-defined concept due to the frequency dependence of the speed of wave-propagation.
Within our framework, it is nonetheless possible to ask questions about black hole superradiance by studying the behaviour of the solutions around turning points, independent of the existence of a universal horizon. 

\section{Overview}

After establishing our general framework, we demonstrate it's usefulness by applying it to the problem of dispersive gravity wave scattering around a draining vortex flow, finding that there are a total of six possible scattering outcomes.
We focus on the deep water regime, an approximation to the full dispersion relation for short wavelengths with respect to the water depth.
As well as being the most relevant regime for existing analogue experiments \cite{torres2017rotational,torres2018application}, this is also the most mathematically consistent treatment of the problem, since the WKB approximation becomes exact in the limit that the wavelength goes to zero.
We find that the transition between different scattering outcomes is delineated in parameter space by four important frequencies.
The first pair are the co- and counter-rotating light-ring ring frequencies, familiar from the stationary orbits of light-rays (or null geodesics) around black holes, whereas the remaining two are related to the strength of dispersion.
In particular, one of the latter is related to a negative energy mode entering the vortex and gives a necessary condition for superradiance, which appears as a simple modification to the non-dispersive result.
However, a careful analysis reveals that this condition is not sufficient to observe amplification at infinity, since the amplified mode can be reflected back into the vortex core by scattering with one of the extra modes in the system.
The possibility of re-scattering is related to the co-rotating light-ring frequency.
This finding constitutes the principle result of this work.

Finally, we provide approximate expressions for the reflection coefficients in each of the scattering scenarios.
In contrast to the behaviour of the reflection coefficient in the non-dispersive case, we find that there can be a frequency range in which the reflection coefficient stays close to unity directly following the range where amplification occurs.
The reason behind this is that dispersion can completely prohibit the propagation of long-wavelength modes in the vortex core, meaning that they are completely reflected.

\section{The system}

The method we present can be applied to a broad class of systems whose linear fluctuations obey the wave equation below.
The dispersive character of the waves will be determined by the physical system under consideration\silke{,} and the scattering of these waves is dictated by the geometry of the set-up.
We now address each of these aspects individually.

\subsection{The wave equation}

Consider a general wave equation in $(N+1)$ dimensions of the form,
\begin{equation} \label{wave_equation}
D_t^2\phi + F(-i\bm{\nabla})\phi = 0,
\end{equation}
where $\phi$ represents the fluctuations and $F$ is an arbitrary function of the gradient operator.
In the context of fluid mechanics, $\mathbf{v}$ corresponds to the velocity field of the background medium and $D_t=\partial_t+\mathbf{v}\cdot\bm{\nabla}$ is the material derivative. 
In general relativity, $\mathbf{v}$ represents the shift vector appearing the in metric when splitting into space and time components \cite{arnowitt1962dynamics}.
This wave equation neglects dissipation but accounts for generic dispersion through the function $F$.
It can be derived as the Euler-Lagrange equation of the following action \cite{coutant2014undulations},
\begin{equation}
\mathcal{S} = \frac{1}{2}\int\left[(D_t\phi)^2 - \phi F(-i\bm\nabla)\phi\right]d^N\mathbf{x}dt.
\end{equation}
Using the action, one can derive conserved currents by appling Noether's theorem for symmetries of the action \cite{schwartz2014quantum}.
For example, for the internal symmetry $\phi\to\phi e^{i\alpha}$ where $\alpha$ is a phase shift, one obtains the conservation equation for the norm current,
\begin{equation}
\partial_t\rho[\phi] + \bm{\nabla}\cdot\mathbf{J}[\phi] = 0,
\end{equation}
where the norm is defined \cite{coutant2016imprint},
\begin{equation} \label{norm}
(\phi,\phi) = \int\rho[\phi]d^2\mathbf{x} = -\int\mathrm{Im}[\phi^*D_t\phi]d^N\mathbf{x}.
\end{equation}
and $\mathbf{J}[\phi]$ is the corresponding current.


\subsection{The dispersion function}

Firstly, one must make a choice for the dispersion function $F$.
Our model example will consist of a body of water at depth $h$ moving with velocity $\mathbf{v}$ in the $(x,y)$ plane (i.e. $N=2$). 
Fluctuations to the water's surface $\delta h$ (known as surface gravity waves) are described by the equation of motion \cite{torres2018waves},
\begin{equation} \label{wave_equationGWs}
D_t^2\phi-ig\bm{\nabla}\cdot\tanh(-ih\bm{\nabla})\phi = 0,
\end{equation}
which is precisely of the form in \eqref{wave_equation}.
Here, $\phi$ is identified with a perturbation of the velocity potential which is related to the free surface fluctuations via,
\begin{equation}
\delta h = -g^{-1}D_t\phi.
\end{equation}
When the wavelength of the fluctuations is much larger than $h$, one may work with a truncation of the hyperbolic tangent function in \eqref{wave_equationGWs} to leading order in it's argument.
This regime, known as shallow water, has the wave equation,
\begin{equation} \label{shallow}
D_t^2\phi - c^2\nabla^2\phi = 0,
\end{equation}
where $c=\sqrt{gh}$ is the shallow water wave speed.
Since all frequencies propagate at this speed, the system is non-dispersive.
Note that \eqref{shallow} is obtained as the low frequency behaviour of a wide variety of systems \cite{barcelo2011analogue} besides that of gravity waves on open channel flows.
All that is required is that the leading term in the Taylor expansion of $F$ is quadratic in $k$.

The wave equation \eqref{shallow} is formally equivalent to the Klein-Gordon equation for a massless scalar field $\phi$,
\begin{equation} \label{KGeq}
\frac{1}{\sqrt{-g}}\partial_\mu\left(\sqrt{-g}g^{\mu\nu}\partial_\nu\phi\right) = 0 \, ,
\end{equation}
which describes how $\phi$ moves through an effective spacetime whose metric is,
\begin{equation}
g_{\mu\nu} = \begin{pmatrix}
-c^2+\mathbf{v}^2 & -\mathbf{v} \\
-\mathbf{v} & \mathbf{I}
\end{pmatrix},
\end{equation}
where $\mathbf{I}$ is the $N\times N$ identity matrix.
The equivalence between \eqref{shallow} and \eqref{KGeq} forms the basis of the analogy between fluid mechanics and general relativity.
As noted above, this limiting behaviour is not unique to gravity waves and as such, fluctuations in a variety of systems can be described in terms of an effective spacetime geometry \cite{barcelo2011analogue}.
In what follows, we will be interested in how this description is modified when dispersive effects are included.

\subsection{Model set-up}

Finally, one must choose the function $\mathbf{v}$ which determines the coordinate dependence of the background.
Our model set-up will be an effectively two dimensional irrotational vortex flow, composed of an inviscid, incompressible fluid.
If the system is axisymmetric and stationary, the general solution to the incompressible and irrotational conditions ($\bm{\nabla}\cdot\mathbf{v}=0$ and $\bm{\nabla}\times\mathbf{v}=0$ respectively) in polar coordinates is,
\begin{equation} \label{DBT}
\mathbf{v} = -\frac{D}{r}\vec{\mathbf{e}}_r + \frac{C}{r}\vec{\mathbf{e}}_\theta.
\end{equation}
Since we will be interested in modelling a draining vortex, we take the drain parameter $D$ to be a positive constant.
The circulation $C$ can in principle take either sign. We choose $C$ to be a positive constant which means the vortex rotates in the direction of increasing $\theta$.
This solution for $\mathbf{v}$ is consistent with the full fluid equations far away from the centre where the water's surface $h$ is approximately uniform.
The flow profile in \eqref{DBT} is known as the draining bathtub vortex (DBT).

In the shallow water regime, this flow profile constitutes the analogue of rotating black hole spacetime, since it exhibits both a horizon and an ergosphere.
The horizon $r_h$ is the boundary of the region inside of which no perturbation can escape to infinity and is given by the condition $|\vec{\mathbf{e}}_r\cdot\mathbf{v}(r_h)|=c$.
The ergosphere $r_e$ is the boundary of the region inside of which no perturbation can move against the flow's rotation which respect to infinity and is given by $|\mathbf{v}(r_e)|=c$.
Solving these two conditions using \eqref{DBT} gives,
\begin{equation}
r_h = \frac{D}{c} \qquad r_e = \frac{\sqrt{C^2+D^2}}{c}.
\end{equation}

\section{The WKB approximation}

\subsection{Homogeneous flow}

When $\mathbf{v}$ is homogeneous, \eqref{wave_equation} admits exact plane wave solutions $\phi\sim \exp(i\mathbf{k}\cdot\mathbf{x}-i\omega t)$, whose frequency $\omega$ and wavevector $\mathbf{k}$ are related through the dispersion relation,
\begin{equation} \label{disp_uniform}
\Omega^2 \equiv (\omega-\mathbf{v}\cdot\mathbf{k})^2 = F(k),
\end{equation}
where $\Omega$ is the intrinsic frequency of the wave in the fluid frame.
The specific $k$ dependence in $F$ will determine the number of solutions $k^j$ to \eqref{disp_uniform}, where $j=1,2,...,M$ where $M$ is the total number of modes.
For $F$ polynomial in $k$, $M$ corresponds to the order of the highest spatial derivative in \eqref{wave_equation}.
Throughout this work, superscript $j$ will indicate that a quantity is associated to a particular $k^j$ mode.

Since \eqref{wave_equation} is second order in time, solutions to the dispersion relation can lie on one of two branches given by,
\begin{equation} \label{branches}
\omega_D^\pm = \mathbf{v}\cdot\mathbf{k}\pm\sqrt{F(k)}.
\end{equation}
The dispersion function $F(k)$ determines the group velocity of the waves via,
\begin{equation} \label{groupvel}
\bm{v}_g = \bm{\nabla}_\mathbf{k}\omega = \mathbf{v}\pm\bm{\nabla}_\mathbf{k}\sqrt{F}.
\end{equation}
This is frequency independent only when $F$ is quadratic in $k$, which corresponds to \eqref{wave_equation} being second order in spatial derivatives.
For any other $k$ dependence, $\bm{v}_g$ becomes frequency dependent and the system is dispersive.

\subsection{Inhomogeneous flow}

When $\mathbf{v}$ is non-uniform, plane waves will no longer be solutions to \eqref{wave_equation}.
However, if the fluctuations vary over a scale $\lambda$ which is much shorter that the scale $L$ over which $\mathbf{v}$ changes, one can define a small parameter $\epsilon=\lambda/L\ll 1$ and write the solution to \eqref{wave_equation} as,
\begin{equation} \label{WKBansatz}
\phi = \mathcal{A}(\mathbf{x},t)\exp\left(\frac{iS(\mathbf{x},t)}{\epsilon}\right),
\end{equation}
where $\mathcal{A}$ and $S$ are the local amplitude and phase respectively.
Inserting \eqref{WKBansatz} into the wave equation \eqref{wave_equation}, the leading contribution in $\epsilon$ gives the Hamilton-Jacobi equation,
\begin{equation} \label{HamJac1}
\left(\partial_t S + \mathbf{v}\cdot\bm{\nabla}S\right)^2 - F(\bm{\nabla} S) = 0.
\end{equation}
This derivation is explained in more detail in \cite{torres2018waves}.
Identifying the frequency and wavevector through,
\begin{equation} \label{Def1}
\omega = -\partial_t S, \qquad \mathbf{k} = \bm{\nabla}S,
\end{equation}
the Hamilton-Jacobi equation is equivalent to the dispersion relation \eqref{disp_uniform} which now gives the local values of $\omega$ and $\mathbf{k}$ when $\mathbf{v}$ is varying.
Since \eqref{HamJac1} is a first order PDE, its solution can be obtained by first splitting into a system of first order ODEs and solving these for the integral (or characteristic) curves.
These characteristics (known as rays in optics and geodesics in general relativity) can be found from an effective Hamiltonian $\mathcal{H}$. Using \eqref{branches}, this can be expressed concisely as,
\begin{equation} \label{Hamiltonian}
\mathcal{H} = -\frac{1}{2}(\omega-\omega_D^+)(\omega-\omega_D^-).
\end{equation}
The characteristics are obtained as the solutions of Hamilton's equations,
\begin{equation} \label{HamiltonsEqs}
\dot{x}^\mu = \frac{\partial\mathcal{H}}{\partial k_\mu}, \qquad \dot{k}_\mu = -\frac{\partial\mathcal{H}}{\partial x^\mu}
\end{equation}
where $x^\mu=(\mathbf{x},t)$, $k_\mu=(\mathbf{k},\omega)$ and the overdot denotes the derivative with respect to $\tau$ which parametrises the curves.
Solving the system of equations \eqref{HamiltonsEqs} gives the coordinates and the conjugate momenta in terms of the parameter $\tau$, i.e. $x^\mu=x^\mu(\tau)$ and $k^\mu=k^\mu(\tau)$.
The phase part of $\phi$ in \eqref{WKBansatz} can then be reconstructed by integrating \eqref{Def1} along the different trajectories.
In addition to \eqref{HamiltonsEqs}, the solutions are also required to satisfy the Hamiltonian constraint,
\begin{equation} \label{onshell}
\mathcal{H}=0,
\end{equation}
which guarantees that they lie on one of the two branches of the dispersion relation \eqref{disp_uniform}.
A solution which satisfies this condition is called \textit{on-shell}, a name borrowed from quantum field theory to describe particles which satisfy the relativistic energy momentum relation \cite{schwartz2014quantum}.

At next to leading order in $\epsilon$, the wave equation gives a transport equation for the amplitude,
\begin{equation} \label{transport}
\partial_t(\Omega\mathcal{A}^2) + \bm{\nabla}\cdot(\bm{v}_g\Omega\mathcal{A}^2) = 0,
\end{equation}
which can be solved for $\mathcal{A}$ using the solutions of the Hamilton-Jacobi equation \eqref{HamJac1}.
This equation describes how the amplitude evolves adiabatically along the characteristics.
As noted earlier, \eqref{transport} fails to account for non-adiabatic exchanges between different modes.
This motivates the development of matching procedures outlined shortly.

\subsection{Stationary systems}

The difficulty of the problem is reduced significantly when $\mathbf{v}$ does not evolve in time, which means that each frequency component evolves independently of the others.
The same is true when the system exhibits some degree of spatial symmetry, for example, if $\mathbf{v}$ is independent of the azimuthal angle $\theta$ as in \eqref{DBT}.
In this case, each of the azimuthal components also evolves independently.
Perturbations can then be decomposed as,
\begin{equation}
\phi(r,\theta,t) = \sum_{m=-\infty}^{+\infty} \frac{\psi(r)}{\sqrt{r}}e^{im\theta-i\omega t}
\end{equation}
where $m$ is the azimuthal number and $\psi$ is the radial mode, i.e. the part of the field containing the $r$ dependence.
The factor of $\sqrt{r}$ is introduced for convenience.
Under these conditions, the wave equation \eqref{wave_equation} becomes an ordinary differential equation in $r$ for $\psi$, which we can solve for using the WKB framework established in the previous section.
In these coordinates, the wavevector has components,
\begin{equation} \label{equation_k}
\mathbf{k} = (p,m/r), \qquad k = \sqrt{\p^2+\frac{m^2}{r^2}}
\end{equation}
where $p$ is the radial wavevector and $k=|\mathbf{k}|$.
The radial WKB modes are given by,
\begin{equation} \label{radial_WKB}
\psi^j = A^j(r) e^{i\int p^j(r) dr}.
\end{equation}
An added benefit of this effectively one dimensional treatment is that $p^j(r)$ can be obtained directly from the dispersion relation \eqref{disp_uniform} for fixed $\omega$ and $m$.
This is equivalent to (but far simpler than) solving Hamilton's equations \eqref{HamiltonsEqs}, since the former is an algebraic problem where as the latter involves differential equations.
The amplitudes $A^j$ are obtained by solving the transport equation \eqref{transport} for each $p^j$.
Using \eqref{disp_uniform}, \eqref{groupvel} and \eqref{Hamiltonian} to write $\mathcal{H}'=\vec{\mathbf{e}}_r\cdot\bm{v}_g\Omega$, where prime denotes derivative with respect to $p$, one finds,
\begin{equation} \label{amp_wkb}
A^j = \alpha^j|\mathcal{H}'(p^j)|^{-\frac{1}{2}}\,
\end{equation}
where $\alpha^j$ is an adiabatically conserved constant of motion.
Using \eqref{radial_WKB}, one can evaluate the norm in \eqref{norm} for the $j^\mathrm{th}$ WKB mode,
\begin{equation} \label{WKBnorm}
\rho[\phi^j] = \Omega(p^j)|A^j|^2.
\end{equation}
In stationary systems, the norm is equivalent to the energy up to a factor of $\omega$ and thus, the sign of the norm and the energy coincide when considering positive frequency mode $\omega>0$.
Hence, a mode with $\Omega(p^j)<0$ carries negative energy.
The corresponding current dictates the direction of energy flow and is given by \cite{richartz2013dispersive,coutant2016imprint},
\begin{equation} \label{energy_current}
\sum_j \mathcal{H}'(p^j)|A^j|^2 = \mathrm{const},
\end{equation}
which is a conserved quantity along $r$.
These last two equations play an essential role in the study of superradiant scattering.

\begin{figure} 
\centering
\includegraphics[width=\linewidth]{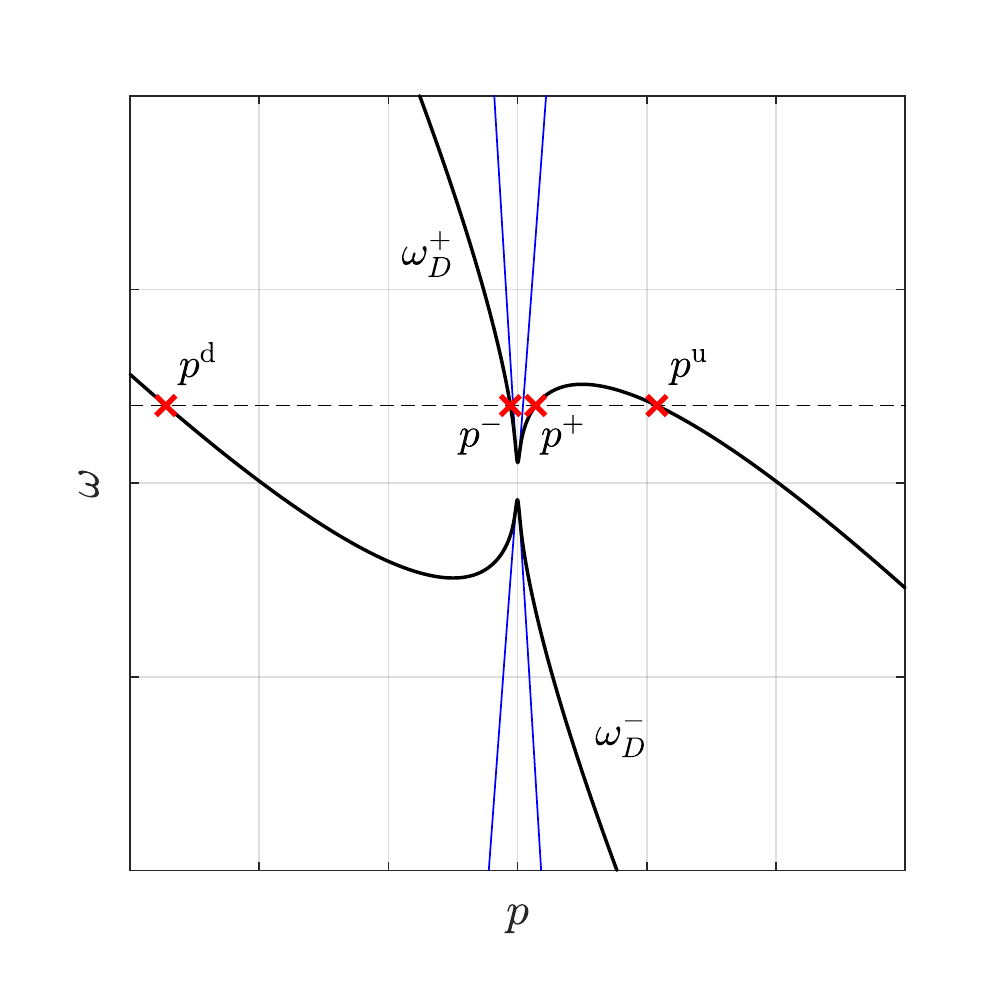}
\caption{The different branches of the dispersion relation for \eqref{branchesGWDBT} at fixed $m$, $C$, $D$, $h$ and $r$ as shown in black.
The skew of the branches is due to the $pD/r$ term in \eqref{branchesGWDBT} and the gap between $\omega_D^\pm$ at $p=0$ results from the $m^2/r^2$ term in $k$, see (\ref{equation_k}).
Both branches are either raised or lowered by the $mv_\theta/r$ term depending on the sign of $m$.
The four $p^j$ correspond to the intersections of a line of fixed $\omega$ with the branches; an example is given of this is given.
Also shown (blue) is the shallow water approximation to the dispersion relation, which only has two solutions and is valid for small $p$.} \label{fig:branches}
\end{figure}

As an example, consider the dispersion relation for gravity waves in the flow field of \eqref{DBT},
\begin{equation} \label{dispDBT}
\left(\omega - \frac{mC}{r^2} + \frac{pD}{r}\right)^2 = gk\tanh(hk).
\end{equation}
The different solutions $p^j$ are given by the intersections of a line of constant $\omega$ with one of the branches of the dispersion relation,
\begin{equation} \label{branchesGWDBT}
\omega_D^\pm = \frac{mC}{r^2} - \frac{pD}{r} \pm \sqrt{gk\tanh(hk)}.
\end{equation}
An example is shown in Fig.~\eqref{fig:branches}.
This particular example has $M=4$ solutions which are labelled $j\in\{\mathrm{d},-,+,\mathrm{u}\}$ in order of increasing $p$.

\subsection{The scattering matrix} \label{sec:scattering_matrix}

The scattering matrix $\mathcal{M}$ is an $M\times M$ matrix which acts on the WKB amplitudes defined in \eqref{amp_wkb} at a point $r_a$ and gives their value at another point $r_b<r_a$,
\begin{equation} \label{scatterMat}
\mathbf{A}_{a,b} = \mathcal{M} \vec{\mathbf{A}}_b,
\end{equation}
where $\mathbf{A}_{a,b}$ is an $M$ component column vector containing all the $A^j_{a,b} = A^j(r_{a,b})$ and $r_{a,b}$ are usually defined to be the edges of the system.
As a matter of convention, we will always order the amplitudes in $\mathbf{A}$ so that the mode with the largest value of $\mathrm{Re}[p^j]$ appears at the top and $\mathrm{Re}[p^j]$ decreases moving down the column vector.

The role of $\mathcal{M}$ is to estimate the amount of mixing between the different modes in the system. 
If $\mathcal{M}$ is a diagonal matrix then each mode evolves independently of the others and no scattering occurs.
When couplings between the modes are included, $\mathcal{M}$ acquires off-diagonal terms which lead to mode-mixing.
As already argued, the dominant contributions to the scattering come from the classical turning points, where neighbouring trajectories coincide and the wave reverses it's direction.
In this work, we focus only on the scattering due to classical turning points, although when mode couplings are computed via other methods (e.g.  \cite{coutant2016imprint,torres2020estimate}) they contribute in form of additional off-diagonal terms in $\mathcal{M}$.

The construction of $\mathcal{M}$ proceeds in the following manner.
\begin{enumerate}
\item Between turning points, the WKB approximation is valid and each mode will evolve according to \eqref{WKBansatz}.
This is encoded the $M\times M$ transport matrix, to be defined later, which transports the solutions adiabatically from one point to another.
\item At the turning points, the modes mix with each other as a result of the non-adiabatic coupling taking place there.
This mixing is computed via an asymptotic matching procedure outlined in the next section, and the result can be expressed as a matrix which transfers the WKB solution across the turning point.
\item Hence, to compute the full scattering matrix, we can successively combine a series of $M\times M$ matrices from point to point as we traverse the system from one side to another.
\end{enumerate}

\subsection{Turning points} \label{sec:tps}

At fixed $\omega$ and $m$, let the location of a turning point be $r_\tp$ with corresponding momentum $p_\tp$.
These two values are obtained by simultaneously solving,
\begin{equation} \label{TPcond}
\mathcal{H}_\tp = 0, \qquad  \mathcal{H}'_\tp = 0,
\end{equation}
where subscript $\tp$ denotes that a quantity is evaluated on the turning point.
The second condition means that the WKB amplitudes in \eqref{amp_wkb} diverge as expected.
The conditions \eqref{TPcond} on the Hamiltonian are equivalent to the following conditions on the dispersion relation,
\begin{equation} 
\omega = {\omega_D^\pm} (r_\tp, p_\tp), \qquad \partial_p\omega_D^\pm \big|_{r_\tp, p_\tp} = 0,
\end{equation}
which means that the turning points are given by the intersections of $\omega=\mathrm{const}$ lines with the extrema of $\omega_D^\pm$ (see e.g. Fig.~\ref{fig:branches}).
When this happens, two of the $p^j$ have the same value, i.e. they interact in the $(r,p)$ plane.
This interaction means that there will be conversion between the two modes and scattering will occur.
Beyond the turning point, the interacting $p^j$ become complex and $\omega=\mathrm{const}$ does not intersect with $\omega_D^\pm$.
Within semiclassical quantum mechanics, this is a forbidden region for classical particles.
In terms of the waves, the solutions to the wave equation will be evanescent and decay spatially.
For the purposes of this section, let the interacting modes be denoted $p^{\IN,\OU}$ when they are real. $p^\IN<p^\OU$ so that $p^\IN$ is to the left of $p^\OU$ when plotted on the dispersion relation.
In the forbidden region, they are labelled $p^\uparrow$ (growing with increasing $r$) and $p^\downarrow$ (decaying with increasing $r$).
These modes only become real and propagating again if they see another turning point.

Before proceeding, we briefly outline the spirit of the calculation.
To find an exact solution around the turning point, one needs to find the local form of the wave equation to linear order in $r-r_\tp$.
Since a turning point involves a two mode interaction, the wave equation will be second order in spatial derivatives and exhibits two independent solutions.
Once obtained, the asymptotic form of these is mapped directly onto the WKB modes close to the turning point and in doing so, one can smoothly connect the different WKB modes either side of $r_{tp}$.
In particular, the relation between the mode amplitudes across the turning point
can be collected into a matrix which transfers the WKB solution from one side of $r_{tp}$ to the other.
The goal of this section will be to find the form of this transfer matrix.

Since the functional form of the exact and WKB solutions are smoothly connected in the matching region, the field and it's first derivative are guaranteed to be continuous, which is the usual requirement of matching procedures for solutions to second order differential equations.
This method implicitly assumes that the matching region is sufficiently close to $r_{tp}$ that the linear expansion of the wave equation is valid but is large enough that the exact solution can reach it's asymptotic value.
Balancing these two requirements yields a validity condition for the procedure (see e.g. \cite{coutant2012black}).
However, we shall soon see that the exact solutions in our case are Airy functions \cite{abramowitz1965handbook}, which rapidly approach their asymptotic value moving away from the turning point.
Furthermore, the argument of the Airy function becomes larger as $|m|$ is increased, therefore one can always find a scenario where the approximation is valid simply by increasing the value of $|m|$.
One final requirement is that when there are multiple turning points, these must be far enough apart that the WKB solutions give a valid approximation in between.
When two turning points become close, one can instead find an exact solution about a saddle point \cite{torres2020estimate}.
We do not explore this here and instead restrict our attention to scenarios where the turning points are far apart.

Now for the analysis.
The local form of the wave equation in the neighbourhood of $r_{tp}$ can be obtained by first expanding the Hamiltonian \eqref{Hamiltonian}.
At fixed $\omega$ and $m$, the Hamiltonian is a function of $r$ and $\p$ only, which at leading order is given by,
\begin{equation} \label{htp0}
\mathcal{H}(r,p) = \partial_r\mathcal{H}_{tp}(r-r_{tp}) + \tfrac{1}{2}\mathcal{H}''_{tp}(\p-\p_{tp})^2,
\end{equation}
where we have used the conditions in \eqref{TPcond}.
For solutions to the dispersion relation, one has $\mathcal{H}=0$.
Promoting $p\to-i\partial_r$, equation \eqref{htp0} may be rewritten as the leading contribution to wave equation,
\begin{equation} \label{Airy1}
-\partial_r^2\psi_{tp} + 2ip_{tp}\partial_r\psi_{tp} + \left[p^2_{tp} + Q(r-r_{tp})\right]\psi_{tp} = 0,
\end{equation}
where $Q = 2\partial_r\mathcal{H}_{tp}/\mathcal{H}''_{tp}$ which is a constant factor determined by the properties of the turning point.
Note that $Q$ increases with $|m|$.
The general solution to \eqref{Airy1} is,
\begin{equation} \label{Airy2}
\psi_{tp} = e^{ip_{tp}r}\left[C_1\mathrm{Ai}\left(s\right) + C_2\mathrm{Bi}\left(s\right)\right],
\end{equation}
where $\mathrm{Ai}(s)$ and $\mathrm{Bi}(s)$ are the two linearly independent solutions of Airy's equation \cite{abramowitz1965handbook}, $C_{1,2}$ are constants and we have defined $s=Q^{1/3}(r-r_{tp})$. Sufficiently far from the turning point, i.e. in the limits $s\to\pm\infty$, these asymptote to,
\begin{equation} \label{AiryAsym}
\begin{split}
\mathrm{Ai}(s) & \underset{-\infty}\sim \frac{1}{2|s|^{1/4}\sqrt{\pi}}\Big(e^{-i\frac{2}{3}(-s)^{3/2}+i\frac{\pi}{4}}+e^{i\frac{2}{3}(-s)^{3/2}-i\frac{\pi}{4}}\Big), \\
& \underset{+\infty}\sim \frac{e^{-\frac{2}{3}s^{3/2}}}{2|s|^{1/4}\sqrt{\pi}}, \\
\mathrm{Bi}(s) & \underset{-\infty}\sim \frac{i}{2|s|^{1/4}\sqrt{\pi}}\Big(e^{i\frac{2}{3}(-s)^{3/2}-i\frac{\pi}{4}}-e^{-i\frac{2}{3}(-s)^{3/2}+i\frac{\pi}{4}}\Big), \\
& \underset{+\infty}\sim \frac{e^{\frac{2}{3}s^{3/2}}}{
|s|^{1/4}\sqrt{\pi}}.\\
\end{split}
\end{equation}
Next, one must find the form of the WKB solutions close to the turning point.
First, solving \eqref{htp0} for $\mathcal{H}=0$ yields the radial wavevector in terms of $s$,
\begin{equation}
p = p_{tp} \pm Q^\frac{1}{3}(-s)^\frac{1}{2}.
\end{equation}
Also using \eqref{htp0} to compute the leading contribution to to the amplitude \eqref{amp_wkb}, the WKB solution becomes,
\begin{equation} \label{wkb_TP}
\psi_\mathrm{WKB} \sim \frac{e^{i\p_\tp r}}{|s|^{1/4}}
 e^{\pm\frac{2}{3}i(-s)^\frac{3}{2}}.
\end{equation}
Consider the scenario where the modes are oscillatory for $s<0$ and evanescent for $s>0$.
The solution either side of the turning point is,
\begin{equation}
\begin{split}
\psi(s<0) = & \ \alpha^\OU\psi^\OU(s)+\alpha^\IN\psi^\IN(s), \\
\psi(s>0) = & \ \alpha^{\downarrow}\psi^{\downarrow}(s)+\alpha^{\uparrow}\psi^{\uparrow}(s).
\end{split}
\end{equation}
where we have defined,
\begin{equation} \label{Airy3}
\begin{split}
\psi^\IN\simeq\frac{e^{-i\frac{2}{3}(-s)^{3/2}}}{2|s|^{1/4}\sqrt{\pi}}, \qquad \psi^\OU\simeq\frac{e^{i\frac{2}{3}(-s)^{3/2}}}{2|s|^{1/4}\sqrt{\pi}}, \\
\psi^{\uparrow}\simeq\frac{e^{\frac{2}{3}s^{3/2}}}{2|s|^{1/4}\sqrt{\pi}}, \qquad \psi^{\downarrow}\simeq\frac{e^{-\frac{2}{3}s^{3/2}}}{2|s|^{1/4}\sqrt{\pi}}.
\end{split}
\end{equation}
By comparing these with $\psi_\mathrm{tp}$ in \eqref{Airy1}, one finds that the amplitudes are related by,
\begin{equation} \label{1tp_cf}
\begin{pmatrix}
\alpha^\OU \\ \alpha^\IN
\end{pmatrix}
= T \begin{pmatrix}
\alpha^{\downarrow} \\ \alpha^{\uparrow}
\end{pmatrix}, \qquad T = e^{\frac{i\pi}{4}}\begin{pmatrix}
1 & -\tfrac{i}{2}\\
-i & \tfrac{1}{2}
\end{pmatrix},
\end{equation}
where $T$ is called the transfer matrix.
Similarly, when the decaying modes are at $s<0$ and the oscillatory modes $s>0$, the solutions above can be used with the transformation $s\to-s$ to show that the coefficients there obey,
\begin{equation} \label{2tp_cf}
\begin{pmatrix}
\alpha^\uparrow \\ \alpha^\downarrow
\end{pmatrix}
= \widetilde{T} \begin{pmatrix}
\tilde{\alpha}^\OU \\ \tilde{\alpha}^\IN
\end{pmatrix}, \qquad \widetilde{T} = e^{\frac{i\pi}{4}}\begin{pmatrix}
\tfrac{1}{2} & -\tfrac{i}{2}\\
-i & 1
\end{pmatrix},
\end{equation}
where $\widetilde{T}$ is the complex inverse of $T$.
Note that the $\uparrow$ is defined as the one which grows in the direction of increasing $r$, hence, the labels on the evanescent modes in \eqref{Airy3} need to be swapped when performing the transformation $s\to-s$.
The result is that the location of the evanescent mode amplitudes in the column vectors differs from \eqref{1tp_cf} to \eqref{2tp_cf} and thus, the transport matrix for evanescent modes should be anti-diagonal (this will be illustrated shortly).
Note also that, as defined, $T$ outputs ($\widetilde{T}$ acts on) a column vector containing the mode with the larger of the two wavevectors at the top.
This fits with the convention outlined in Section~\ref{sec:scattering_matrix}.

Many of the scattering scenarios considered in this work include multiple turning points.
Let two such turning points be denoted $r_a$ and $r_b$, with evanescent modes $p^\uparrow$ and $p^\downarrow$ in the region $r_a<r<r_b$.
It will prove extreme useful to define a matrix $\mathcal{N}_{ab}$ which relates the amplitudes ($A^j_b$) of oscillatory modes in the region $r>r_b$ to those ($A^j_b$) at $r<r_a$.
First, we define the shift factor $\SH^j_{ab}$ which adiabatically translates the WKB mode $\psi^j$ from $r_b$ to $r_a$,
\begin{equation} \label{WKBoperator}
\SH^j_{ab} = \sqrt{\left|\frac{\mathcal{H}'_b(p^j)}{\mathcal{H}'_a(p^j)}\right|}\exp\left(-i\int^{r_b}_{r_a} p^j dr \right).
\end{equation}
Note that these functions are scalars and not tensors; the lower indices indicate that the function is applied at $r_b$ and returns an object at $r_a$.
The mode amplitudes are then related by applying the transfer and transport matrices, where the latter is anti-diagonal in the forbidden region,
\begin{equation} \label{LocalScatter}
\begin{pmatrix}
A_a^\OU \\ A_a^\IN
\end{pmatrix} = T \begin{pmatrix}
0 & \SH_{ab}^\downarrow \\ \SH_{ab}^\uparrow & 0
\end{pmatrix} \widetilde{T} \begin{pmatrix}
A_b^\OU \\ A_b^\IN
\end{pmatrix}.
\end{equation}
In the forbidden region, the radial wavevectors satisfy $\mathrm{Re}[p^\uparrow]=\mathrm{Re}[p^\downarrow]$, $\mathrm{Im}[p^\uparrow]=-\mathrm{Im}[p^\downarrow]<0$, and one also has $|\mathcal{H}'(p^\uparrow)|=|\mathcal{H}'(p^\downarrow)|$.
Using these relations, \eqref{LocalScatter} can be rewritten,
\begin{equation}
\begin{pmatrix}
A_a^\OU \\ A_a^\IN
\end{pmatrix} = \SH_{ab}^\downarrow \mathcal{N}_{ab} \begin{pmatrix}
A_b^\OU \\ A_b^\IN
\end{pmatrix},
\end{equation}
where we have defined,
\begin{equation} \label{LocalScatter2}
\begin{split}
\mathcal{N}_{ab} = & \ \begin{pmatrix}
1+\tfrac{1}{4}f_{ab}^2 & i\left(1-\tfrac{1}{4}f_{ab}^2\right) \\ -i\left(1-\tfrac{1}{4}f_{ab}^2\right) & 1+\tfrac{1}{4}f_{ab}^2
\end{pmatrix}, \\
f_{ab} = & \ \exp\left(-\int^{r_b}_{r_a}\mathrm{Im}[p^\downarrow]dr\right).
\end{split}
\end{equation}
The matrix $\mathcal{N}_{ab}$ will serve as the main tool in computing scattering coefficients in this work.

Note, the naive application of this formula to scenarios where $r_b-r_a$ is smaller then the local wavelength (or decay length) will yield erroneous results.
The reason is that the linear expansion in \eqref{htp0} is not valid since the value of $\partial_r\mathcal{H}_\tp$ becomes very small.
In this case, a quadratic expansion of the Hamiltonian is more appropriate, see e.g. \cite{torres2020estimate}.
We do not explore this here,  but note that the method results again in a matrix like $\mathcal{N}_{ab}$, albeit with different components, converting between oscillatory WKB solutions.
This can be easily incorporated into our framework simply be changing the components of the matrix $\mathcal{N}_{ab}$.

\section{Shallow water}

As a first example, we consider scattering in shallow water where $hk\ll 1$.
The dispersion function in \eqref{dispDBT} in this limit reduces to,
\begin{equation} \label{shallowF}
F(k) = c^2k^2,
\end{equation}
where the wave speed is defined $c=\sqrt{gh}$.
This choice of dispersion function corresponds to the blue lines in Fig.~\ref{fig:branches}, which approximate the exact dispersion relation at small $p$ values.

\begin{figure*} 
\centering
\includegraphics{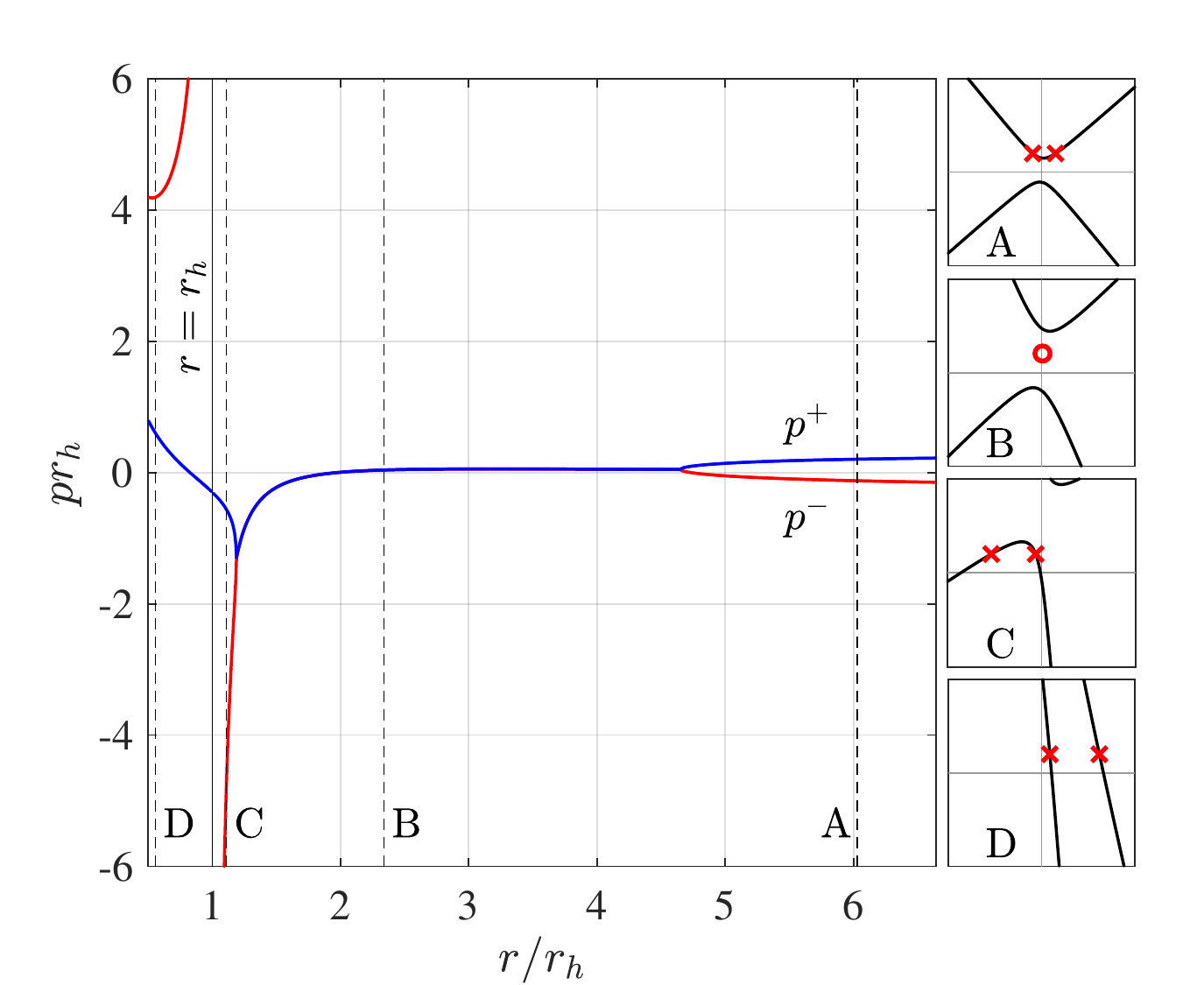}
\caption{An example of $p^\pm(r)$ in \eqref{shallow_sols} with the parameters $C/D=1$, $\omega D/c^2=6\times10^4$ and $m=1$.
Only the real part of $p$ is displayed.
The figure shows two modes propagating on the far right which then meet at a turning point.
The modes then remain evanescent until a second turning point is reached at smaller $r$.
Since $\tilde{\omega}_h<0$ for these parameters, the $+$ mode crosses the horizon whereas the $-$ mode diverges there.
The location of the two modes on the dispersion relation at different values of $r$ are displayed in panels A to D.
The different panels correspond to dashed black lines on the main figure with the same label.
In the panels, red crosses indicate $p\in\mathbb{R}$ whereas the red circle indicates $p\in\mathbb{C}$.} \label{fig:modes_shallow}
\end{figure*}

The shallow water regime possesses a number of attractive features that simplify the analysis significantly.
Firstly, all waves (irrespective of frequency) propagate at the same speed $c$; the system is non-dispersive.
Secondly, the dispersion relation is quadratic in $p$ and only has two solutions: these are the $p^\pm$ modes.
Thirdly, the equation governing the radial trajectories of the modes admits a rewriting in terms of an effective potential,
\begin{equation} \label{potential}
V = -\tilde{\omega}^2 + \left(c^2-v_r^2\right)m^2/r^2,
\end{equation}
where the frequency in the rotating frame is defined,
\begin{equation}
\tilde{\omega} = \omega-mv_\theta/r.
\end{equation}
These can be used to concisely express the exact solution for both $p$ modes,
\begin{equation} \label{shallow_sols}
p_\pm = \frac{-\tilde{\omega}v_r\pm c \sqrt{-V}}{c^2-v_r^2}.
\end{equation}
The solutions become evanescent when $V$ is positive and are denoted $p^\uparrow$ and $p^\downarrow$ as in the previous section. 
An example of $p^\pm(r)$ is plotted in Fig.~\ref{fig:modes_shallow}.
Finally, the solution for $p$ can be used to show that the spatial dependence in the amplitude \eqref{amp_wkb} for both modes is given by,
\begin{equation} \label{Hprime_shallow}
|H'| = |\sqrt{-V}|.
\end{equation}
Since the turning points of $\mathcal{H}$ correspond to the zeros of $V$, scattering can be understood simply in terms of the effective potential \eqref{potential}.
This has a maximum of two zeros.
When these zeros are far apart, the method outlined in Section \ref{sec:tps} for relating the mode amplitudes is applicable.

\subsection{Superradiance} \label{sec:super}

Superradiance occurs when the system absorbs a negative energy mode.
Equivalently, the energy current of a mode propagating into the centre of the vortex needs to be positive.
Such a mode extracts energy from the system, and consequently, the reflected mode at infinity is amplified.
To see this, we can compare the conserved current in \eqref{energy_current} on the horizon and at infinity.
The asymptotic form of the modes in \eqref{radial_WKB} is,
\begin{equation} \label{shallow_asymp}
\begin{split}
\psi(r=r_h) = & \ A^-_h e^{-i\tilde{\omega}_hr_*} \\
\psi(r\to\infty) = & \ A^-_\infty e^{-i\omega r/c} + A^+_\infty e^{i\omega r/c},
\end{split}
\end{equation}
where subscript $h$ denotes that a quantity is evaluated at $r_h$, and $r_*$ is a function of $r$ that goes to $-\infty$ on the horizon \cite{dolan2012resonances}.
On the horizon, the absorbing boundary condition has been used to discard the mode whose radial group velocity is directed toward large $r$.
The mode which is transmitted to smaller $r$ corresponds to the $-$ mode for $\tilde{\omega}_h>0$ and the $+$ mode for $\tilde{\omega}_h<0$.
Evaluating the energy current at these locations gives,
\begin{equation} \label{shallow_Ecurr}
-\tilde{\omega}_h|A_h^\pm|^2 = -\omega|A_\infty^-|^2 + \omega|A_\infty^+|^2.
\end{equation}
The reflection coefficient is defined as the ratio of the energy current of $+$ and $-$ modes at infinity.
However, since the prefactor $\mathcal{H'}$ in the energy current is the same for both $+$ and $-$ modes, the scattering coefficients can be defined as the ratio of the amplitudes,
\begin{equation} \label{shallow_coefs}
\mathcal{R} = \frac{A_\infty^+}{A_\infty^-}, \qquad \mathcal{T} = \frac{A_h^\pm}{A_\infty^-},
\end{equation}
where $\mathcal{R}$ is the reflection coefficient and $\mathcal{T}$ is the transmission coefficient.
Inserting these definitions into \eqref{shallow_Ecurr} gives,
\begin{equation}
|\mathcal{R}|^2+\frac{\tilde{\omega}_h}{\omega}|\mathcal{T}|^2 = 1.
\end{equation}
From this expression, we clearly see that the reflected mode is amplified ($\mathcal{R}>1$) when $\tilde{\omega}_h<0$ is satisfied.
Hence, the condition for superradiance in the shallow water regime is,
\begin{equation} \label{SR_cond1}
\omega < \frac{mC}{r_h^2}.
\end{equation}
Note, although we have used WKB solutions here, this condition can also be derived from the full conserved current of \eqref{wave_equation} and hence is exact.
It will be the challenge of a moment to find the equivalent condition in the deep water regime.

\subsection{Reflection coefficient} \label{sec:shal_refl}

\begin{figure} 
\centering
\includegraphics[width=\linewidth]{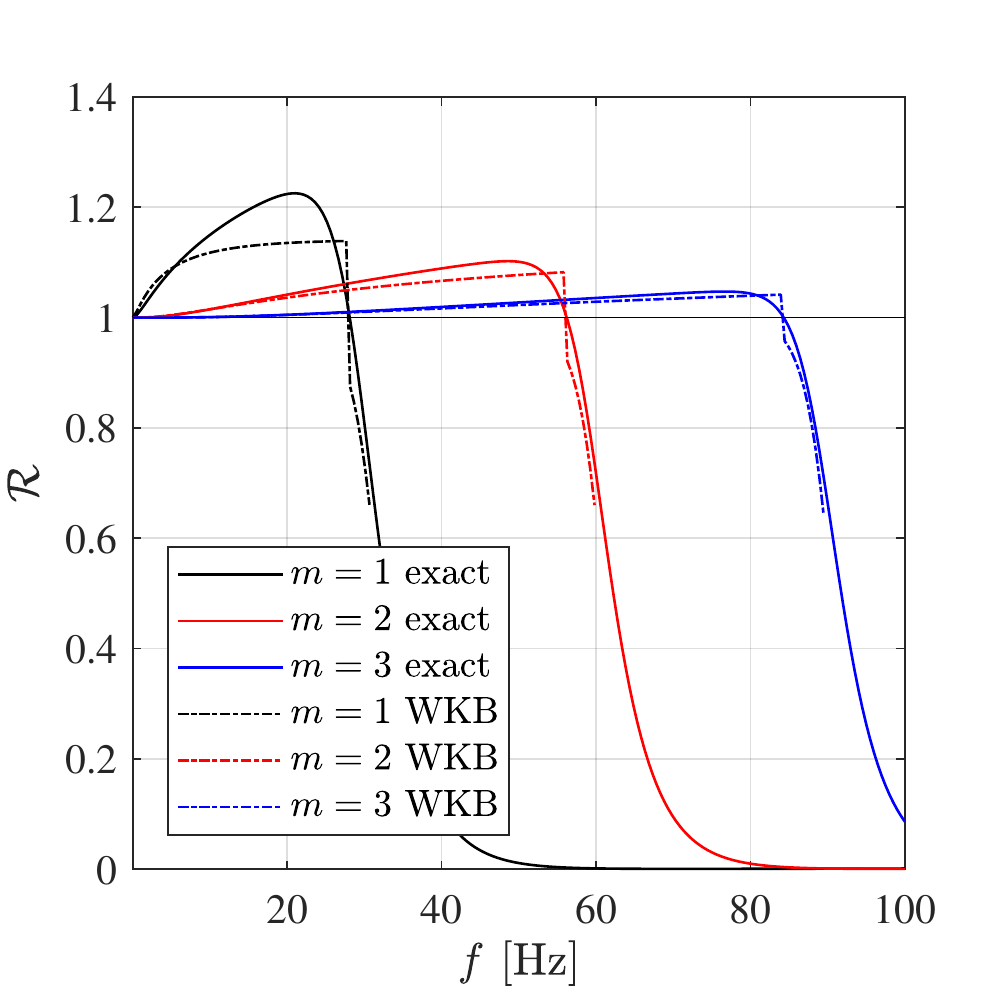}
\caption{Reflection coefficients in shallow water for the flow parameters $C=1.4\times 10^{-2}~\mathrm{m^2/s}$, $D=-7\times 10^{-3}~\mathrm{m^2/s}$, $h=6.3~\mathrm{cm}$.
The solid lines are the exact coefficients computed from numerical simulation.
The broken lines represent WKB prediction in \eqref{ReflWKB}.
Agreement with exact coefficients improves as $m$ is increased.
The dashed lines terminate abruptly since there are no real turning points at higher frequencies, which is a requirement of the approximation used here.
} \label{fig:Shallow}
\end{figure}

When the effective potential has two real zeros $r_1$ and $r_2$ which are far apart, the scattering coefficients are related by application of: a) the diagonal transport matrices in the two classically allowed regions, and b), the matrix $\mathcal{N}_{12}$ defined in \eqref{LocalScatter2} across the forbidden region.
For $\tilde{\omega}_h>0$, we have
\begin{equation} \label{scattering_shallow}
\begin{pmatrix}
0 \\ \mathcal{T}
\end{pmatrix} = \begin{pmatrix}
\SH_{h1}^+ & 0 \\ 0 & \SH_{h1}^-
\end{pmatrix} \SH_{12}^\downarrow\mathcal{N}_{12} \begin{pmatrix}
\SH_{2\infty}^+ & 0 \\ 0 & \SH_{2\infty}^-
\end{pmatrix}  \begin{pmatrix}
\mathcal{R} \\ 1
\end{pmatrix},
\end{equation}
whereas for $\tilde{\omega}_h<0$, $\mathcal{T}$ and 0 swap positions in the column vector since the mode which diverges on the horizon changes.
Dropping the WKB phase factors, which don't affect the magnitude of the scattering coefficients, \eqref{scattering_shallow} becomes,
\begin{equation}
\begin{pmatrix}
0 \\ \mathcal{T}
\end{pmatrix} = \left|\frac{\omega}{\tilde{\omega}_h}\right| \mathcal{N}_{12} \begin{pmatrix}
\mathcal{R} \\ 1
\end{pmatrix}
\end{equation}
Using \eqref{LocalScatter2} and solving for $\mathcal{R}$, one arrives at the following expression,
\begin{equation} \label{ReflWKB}
\mathcal{R} = e^{-\frac{i\pi}{2}}\left(\frac{1-f_{12}^2/4}{1+f_{12}^2/4}\right)^{\mathrm{sgn}(\tilde{\omega}_h)},
\end{equation}
which is greater than $1$ for $\tilde{\omega}_h<0$ as expected.
In Fig.~\eqref{fig:Shallow}, the predictions of \eqref{ReflWKB} are shown to be in good agreement with direct numerical simulation of the shallow water wave equation (detailed in Appendix A).
Agreement improves as $m$ is increased, as one would expect from the WKB approximation. 

The expression in \eqref{ReflWKB} also reveals clearly the asymptotic behaviour of the reflection coefficient.
In particular, since the integral in $f_{12}$ decreases with $\omega$, the maximum value of $\mathcal{R}$ will always occur just below $\omega=m\Omega_h$ within this approximation.
Furthermore, since the integral increases with $m$, this maximum value will decrease exponentially with increasing $m$.
Since the WKB approximation improves as $m$ gets larger, this must also be the asymptotic behaviour exhibited by the exact solutions.


\section{Deep water} \label{sec:deep}

\begin{figure*} 
\centering
\includegraphics{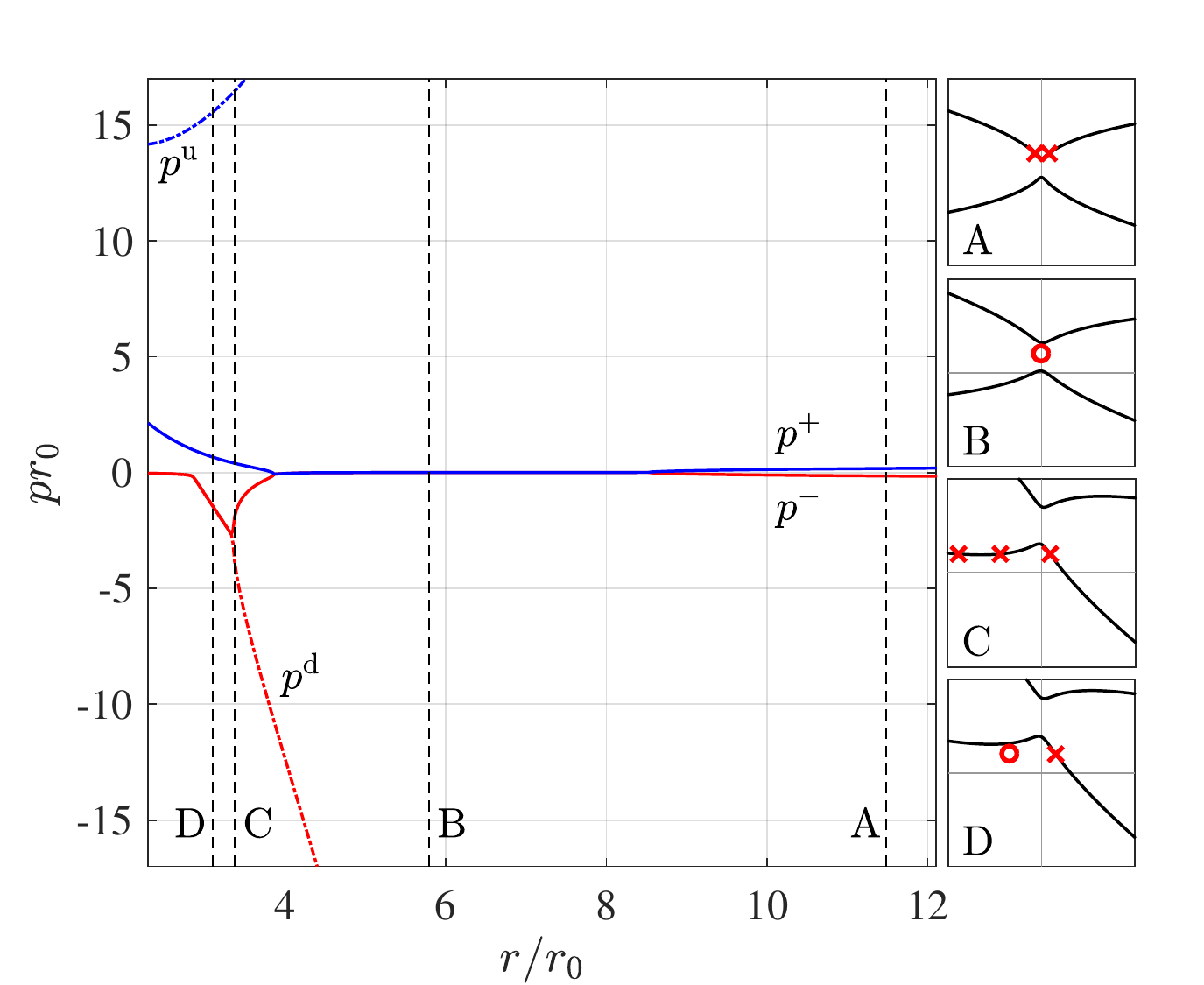}
\caption{An example of 4 solutions $p^j(r)$ to the dispersion relation in \eqref{dispDBT} with the parameters $C/D=16$, $h/r_0=15$, $\omega t_0 = 0.54$, $m=1$.
The deep water solutions only differ notably for the two evanescent modes approaching $r=0$, whose real part diverges at the origin tracking the behaviour of the two real modes there.
This difference is not relevant for our discussion.
At large $r$, the values of $p^\mathrm{u,d}$ become much larger than those of $p^\pm$ and cannot be seen on the plot.
In this case, the evolution of $p^\pm$ proceeds similarly to that described in the shallow water case in Fig.~\ref{fig:modes_shallow} except at small $r$, the horizon is replaced by a turning point where the $-$ mode is converted into the d mode.} \label{fig:modes_deep}
\end{figure*}

In the deep water regime $hk\gg 1$, the dispersion function \eqref{dispDBT} is approximated by,
\begin{equation} \label{DWdisp}
F(k) = g|k|,
\end{equation}
where the modulus is understood to only cancel the overall sign.
This choice of dispersion function is qualitatively indistinguishable from the solid black curves in Fig.~\eqref{fig:branches} except for very small $p$.
As such, \eqref{DWdisp} gives a good approximation to the exact dispersion function over a much wider range in $p$ than the shallow water approximation.
In deep water, the motion of fluctuations occurs predominantly close to the free surface and consequently, the problem is independent of the water depth $h$.
To lighten the notation, we define the characteristic length and time scales,
\begin{equation}
r_0 = \left(\frac{D^2}{g}\right)^\frac{1}{3}, \qquad t_0 = \left(\frac{D}{g^2}\right)^\frac{1}{3}.
\end{equation}
In this section, we will work with a dimensionless rescaling of \eqref{dispDBT} by these parameters (unless otherwise stated) which amounts to setting $D=g=1$.

The absence of the hyperbolic tangent function means that the equations for $p$ will be polynomial rather than transcendental, which makes the analysis analytically tractable.
In the shallow water case, the exact solutions for $p$ took on the closed form expressions in \eqref{shallow_sols}.
Since the equation for $p$ is quartic in the deep water regime, closed form solutions also exist, but their form is not sufficiently enlightening to be worth writing down.
Thus, to find the four values of $p^j$, we solve the quartic equation numerically.
Once these solutions are obtained, our analysis proceeds analytically.

As an example, the four $p^j(r)$ are plotted for specific flow parameters in Fig.~\ref{fig:modes_deep}.
Comparing with the shallow water modes in Fig.~\ref{fig:modes_shallow}, the main difference is that there are now four modes present at all radii as opposed to just two and the horizon has been replaced by a turning point.

In general, the dispersion relation in deep water exhibits either one, three or five real turning points, denoted $r_a$ with $a\in\{1,2,3,4,5\}$ in order of increasing size.
From Fig.~\ref{fig:branches}, we can see that at large enough $r$, all four of the $p^j(r)$ are real and propagating.
Since the number of real turning points is odd, there will always be two real modes and two evanescent modes approaching the centre of the vortex.
Which two modes are real depends on the values of $\omega$, $m$ and $C$.
The turning point $r_1$ is the analogue of the shallow water horizon when a long wavelength $+$ or $-$ mode becomes compressed at small $r$, converts into a short wavelength u or d mode and reverses it's direction.
This conversion between short and long wavelength modes is the key feature of dispersive systems that is absent in shallow water.

\subsection{Scattering types}

\begin{figure*} 
\centering
\includegraphics[width=\linewidth]{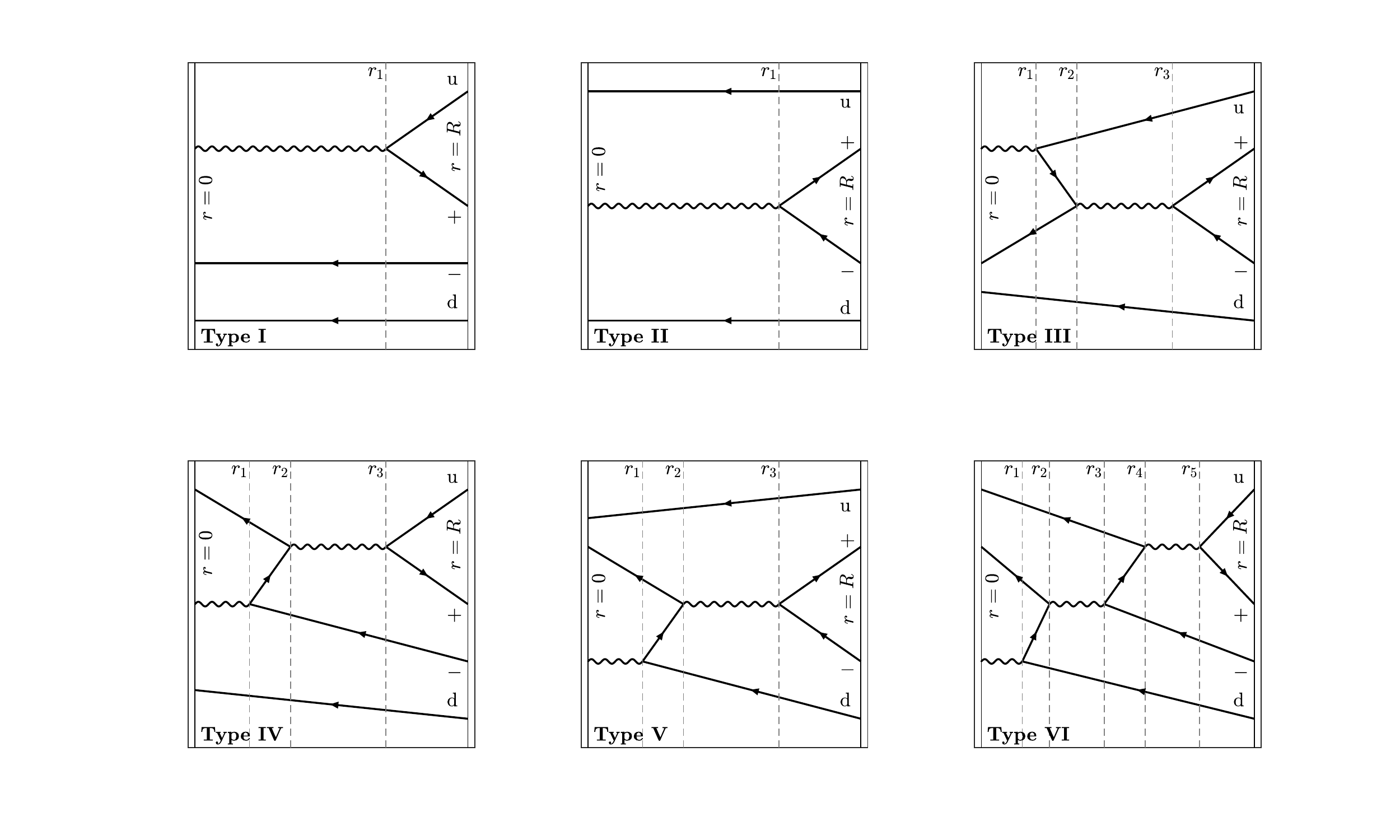}
\caption{The different possible outcomes of a scattering process schematically illustrated as Feynman diagrams.
These diagrams show the different mode interactions that occur in the $r,p$ plots, e.g. Fig.~\ref{fig:modes_deep} corresponds to the type V diagram here.
Moving in the direction of decreasing $r$, ``particles" are in-going waves and ``anti-particles"  are out-going modes.
Wavy lines, or ``virtual-particles" represent evanescent modes.
Vertices correspond to turning points.
} \label{fig:Feynmann_super}
\end{figure*}

The possibility of interactions between long and short wavelength modes greatly enriches the possible outcomes of a scattering event.
For the velocity profiles \eqref{DBT}, the different outcomes can be grouped into 6 different classes, each determined by the number and location of real turning points on the dispersion relation.
The 6 possibilities are illustrated in the form of Feynman diagrams in the $(r,p)$ plane in Fig.~\ref{fig:Feynmann_super}.
Pursuing this analogy with particle physics, one can identify in-going modes with particles and out-going modes with anti-particles.
Evanescent modes are analogous to virtual particles.
Shortly, we assign rules to these diagrams to facilitate the computation of the reflection coefficient, which can be in general a lengthy procedure.

\begin{figure*} 
\centering
\includegraphics[width=\linewidth]{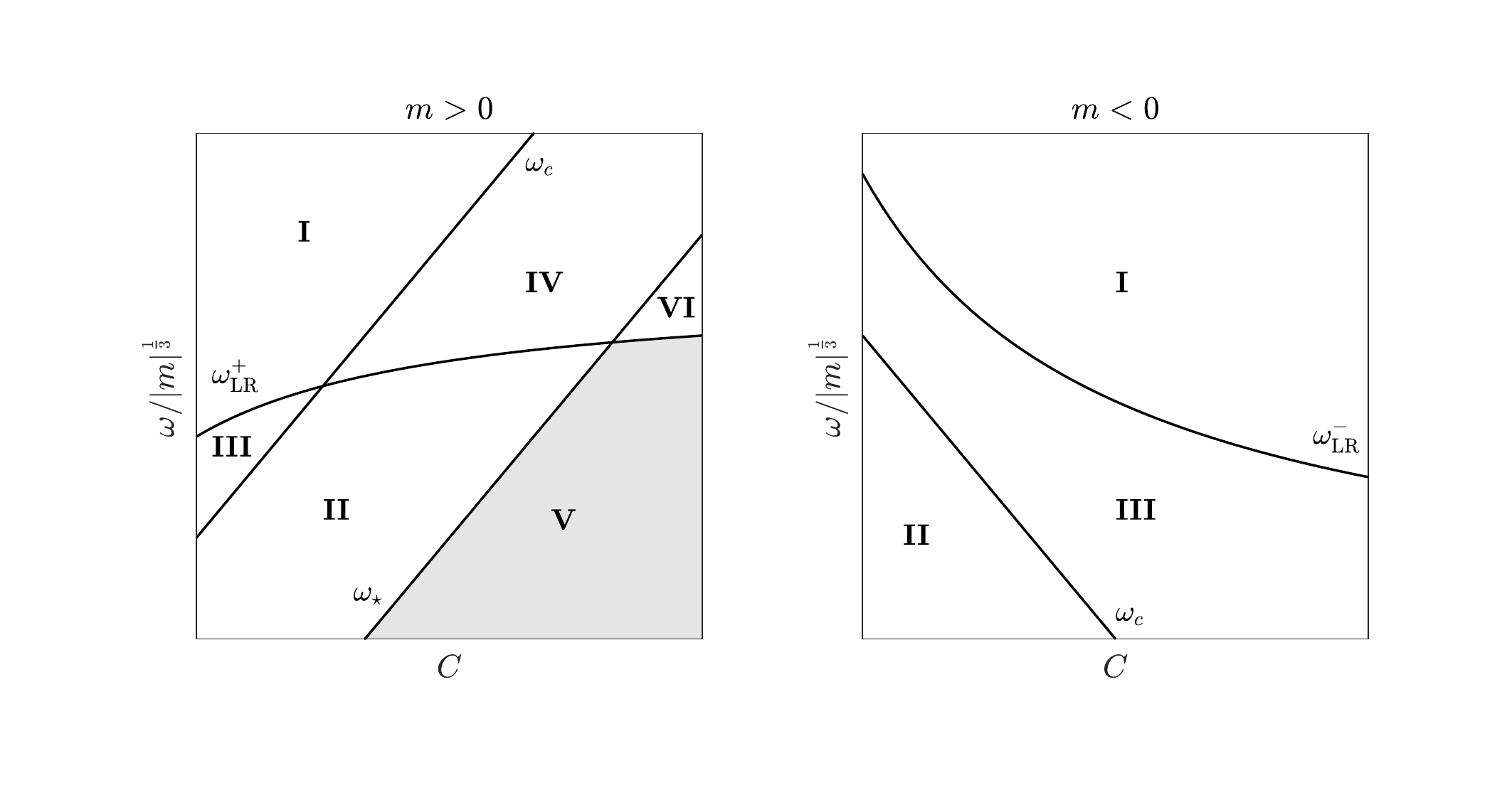}
\caption{The parameter space is split, by the four frequencies in \eqref{LightRings} and \eqref{CritFreqs}, into regions where the different scattering processes in Fig.~\ref{fig:Feynmann_super} occur.
The bottom-left corner in the plots is the point $\omega=0$, $C=0$.
A schematic illustration is provided since certain regions (in particular, the one labelled III for $m>0$) are difficult to resolve when plotting the numerical values of the curves.
There are two distinct structures to the space depending on the sign of $m$.
For $m=0$, only type I scattering occurs.
Amplification is guaranteed in the (grey) region where \eqref{SR_cond2} is satisfied.
This is condition is sufficient for amplification but not always necessary. 
In particular, amplification may also occur just beyond the $\omega_\mathrm{LR}^+$ curve where exponential suppression of the reflected mode is only minor.
} \label{fig:ParamSpace_Types}
\end{figure*}

For $m=0$, there is single real turning point given in dimensionless variables by,
\begin{equation} \label{TPm0}
r_1^{m=0} = 4\omega,
\end{equation}
and only type I scattering occurs.
For $m\neq 0$, there can be up to 5 turning points.
These are solved for numerically in specific examples later on.
There are four important frequencies which determine the number of real turning points and therefore distinguish the different scattering types;
these are the light-ring frequencies and two new frequencies which we call the upper and lower critical frequencies for reasons which will shortly become apparent.
These four frequencies divide up the parameter space into different scattering regions and are schematically illustrated as functions of $C$ in Fig.~\ref{fig:ParamSpace_Types}.

The light-rings $r_\mathrm{LR}$ are the stationary orbits of $\mathcal{H}$ and are given by,
\begin{equation}
\mathcal{H}_\mathrm{LR} = 0, \qquad \mathcal{H}'_\mathrm{LR}=0, \qquad \partial_r\mathcal{H}_\mathrm{LR}=0.
\end{equation}
These conditions can be solved at fixed $m$ and $C$ for the triplet ($r^\pm_\mathrm{LR},p^\pm_\mathrm{LR},\omega^\pm_\mathrm{LR}$) where the $\pm$ sign corresponds to the sign of $m$.
These were previously worked out for the deep water case in \cite{torres2018waves}; in particular, the light-ring frequencies are,
\begin{equation} \label{LightRings}
\begin{split}
\omega^\pm_\mathrm{LR}(m) = & \ \frac{3}{8}\left(\frac{4}{\sqrt{B_\pm^2-1}}\right)^\frac{1}{3} |m|^\frac{1}{3}, \\
B_\pm = & \ \left[2(C^2+1)\mp2C\sqrt{C^2+1}\right]^\frac{1}{2}.
\end{split}
\end{equation}
As noted in \cite{torres2018waves}, given certain conditions these frequencies are related to the quasinormal modes of the system.
For the present purposes, they correspond to scenarios in which two modes interact at a single point in the $(r,p)$ plane before departing.
Hence, the light-ring frequency forms a boundary between two distinct regions of parameter space: in one region, the two modes interact and in the other region they are decoupled.

The second pair of important frequencies are derived from the following consideration.
As $r$ is decreased, the branches of the dispersion relation (e.g. in Fig.~\ref{fig:branches}) become increasingly skewed by the linear in $p$-term in equation \eqref{dispDBT}.
Eventually, the skew becomes sufficiently significant that the extrema of $\omega_D^\pm$ disappear.
Just before this happens, the two pairs of turning points (one on each branch) merge to become inflection points.
Let these inflection points be located at $p_c$ (upper branch) and $p_\star$ (lower branch) respectively, and the values of $\omega$ there are $\omega_c$ and $\omega_\star$.
These critical frequencies play an important role in determining which modes propagate in the vortex core.
On the upper branch, the u mode is real approaching the origin for $\omega<\omega_c$ whereas above $\omega_c$, the $-$ mode is real.
On the lower branch, the d mode is real for $\omega>\omega_\star$ whereas below $\omega_c$, the $+$ mode is real.
Due to the symmetry of the dispersion relation, the following relations are true:
$r_c=r_\star$, $p_c=-p_\star$, $\omega_c(m<0)=-\omega_\star(m>0)$ and $\omega_\star(m<0)=-\omega_c(m>0)$.
Note that since $\omega_\star$ concerns the inflection point on the lower branch, this frequency plays no role for positive frequency modes with $m<0$.
The reason for this is that as $r$ is decreased, the $mC/r^2$ term in \eqref{dispDBT} pushes $\omega_D^-$ to increasingly lower $\omega$.
However, $\omega_\star$ is still important for positive frequency modes with $m>0$, since in this case, the $mC/r^2$ can raise the $\omega_D^-$ branch to positive frequencies. 
These observations are summarised in the parameter space plots of Fig.~\ref{fig:ParamSpace_Types}.

The condition for the inflection points is equivalent to following conditions on the Hamiltonian,
\begin{equation}
\mathcal{H}_{c,\star} = 0, \qquad \mathcal{H}'_{c,\star}=0, \qquad \mathcal{H}''_{c,\star}=0,
\end{equation}
which are solved at fixed $m$ and $C$ for the triplets ($r_c,p_c,\omega_c$) and ($r_\star,p_\star,\omega_\star$).
In the deep water regime, the upper and lower critical frequencies are given by,
\begin{equation} \label{CritFreqs}
\omega_\star = \frac{1}{2^\frac{2}{3}3}\left(C-2^\frac{3}{2}\right)|m|^\frac{1}{3}, \quad \omega_c = \frac{1}{2^\frac{2}{3}3}\left(C+2^\frac{3}{2}\right)|m|^\frac{1}{3} .
\end{equation}
Note that for $C<2^\frac{3}{2}$, $\omega_\star$ becomes negative which means that at low rotation, positive frequency, long wavelength modes cannot propagate in the vortex core approaching $\omega=0$.
This phenomenon is a direct consequence of the deep water dispersion relation which has no analogue in the non-dispersive case.
Another consequence is the $m^\frac{1}{3}$ dependence, as was also noted in \cite{torres2018waves}.


\subsection{Superradiance condition}

Proceeding in the same fashion as in Section~\ref{sec:super}, the existence of superradiance is derived by analysing the conserved current \eqref{energy_current}.
Evaluating this at the innermost turning point $r_1$ and some radius far from the centre (say $r=R$), this gives,
\begin{widetext}
\begin{equation} \label{energy_current2}
\begin{split}
\mathcal{H}'_1(p^\mathrm{d}_1) |A^\mathrm{d}_1|^2 + \mathcal{H}'_1(p^\mathrm{u}_1)|A^\mathrm{u}_1|^2 + \mathcal{H}'_1(p^+_1)|A^+_1|^2 & \ + \mathcal{H}'_1(p^-_1)|A^-_1|^2 \\
= & \ \mathcal{H}'_R(p^\mathrm{d}_R)|A^\mathrm{d}_R|^2 + \mathcal{H}'_R(p^\mathrm{u}_R)|A^\mathrm{u}_R|^2 + \mathcal{H}'_R(p^+_R)|A^+_R|^2 + \mathcal{H}'_R(p^-_R)|A^-_R|^2.
\end{split}
\end{equation}
\end{widetext}
Application of the matrix $\widetilde{T}$ in \eqref{2tp_cf} at $r_1$ reveals that the two interacting modes there (say $\IN$ and $\OU$) satisfy $A^\OU_1=e^{-i\pi/2}A^\IN_1$, as both modes must decay towards the origin.
Since $\mathcal{H}'(p^\IN)=-\mathcal{H}'(p^\OU)$ approaching $r_1$, the contributions of these modes cancel one-another.
Physically, this means that neither mode may carry energy beyond $r_1$.
The allowed pairings ($\IN,\OU$) are ($+$,u) for types I and III, ($-,+$) for types II and IV and (d,$-$) for types V and VI.
The remaining modes (say $\Lambda$ and P) propagate into the centre.

The reflection coefficient $\mathcal{R}$ is defined as, 
\begin{equation} \label{R_coeff}
\mathcal{R} = \left|\frac{\mathcal{H}'_R(p^+_R)}{\mathcal{H}'_R(p^-_R)}\right|^\frac{1}{2}\frac{A^+_R}{A^-_R}.
\end{equation}
In contrast to the shallow water definition, the factors of $\mathcal{H}'$ need to be included since these are not equal for the two modes.
One also needs to specify how much additional energy is carried by the in-going short wavelength modes.
This information is contained in the coefficients,
\begin{equation} \label{I_coeff}
\mathcal{I}^\mathrm{u,d} = \left|\frac{\mathcal{H}'_R(p^\mathrm{u,d}_R)}{\mathcal{H}'_R(p^-_R)}\right|^\frac{1}{2}\frac{A^\mathrm{u,d}_R}{A^-_R}.
\end{equation}
Lastly, the transmission coefficients are,
\begin{equation}
\mathcal{T}^{\Lambda,\mathrm{P}} = \left|\frac{\mathcal{H}'_1(p^{\Lambda,\mathrm{P}}_1)}{\mathcal{H}'_R(p^-_R)}\right|^\frac{1}{2}\frac{A^{\Lambda,\mathrm{P}}_1}{A^-_R}.
\end{equation}
Inserting these definitions into \eqref{energy_current2} gives,
\begin{equation} \label{energy_current4}
|\mathcal{R}|^2 \mp^\Lambda |\mathcal{T}^\Lambda|^2 \mp^\mathrm{P} |\mathcal{T}^\mathrm{P}|^2 = 1 + |\mathcal{I}^\mathrm{u}|^2 - |\mathcal{I}^\mathrm{d}|^2,
\end{equation}
where $\mp^{\Lambda,\mathrm{P}}=-\mathrm{sgn}(\mathcal{H}'_1(p^{\Lambda,\mathrm{P}}_1))$.
In this rewriting of \eqref{energy_current2}, all of the input terms are on the right hand side, whereas the output terms are on the left.
The first thing to notice is that $\mathcal{R}=1$ does not correspond to perfect reflection if $\mathcal{I}^\mathrm{u,d}$ are non-zero and instead, we are looking for $|\mathcal{R}|^2$ to be greater than the sum of the terms on the right hand side of \eqref{energy_current4}.
For this to occur, one of the $\mathcal{T}$ terms must contribute negatively to the left hand side, which happens if $\mathcal{H}'_1(p^{\Lambda,\mathrm{P}}_1)>0$ for one of the two modes.
The scenarios in which this is satisfied have either $\Lambda=\mathrm{d}$ or $\Lambda=+$.
However, the d mode has a negative norm at infinity and is therefore non-physical.
The only physical possibility is $\Lambda=+$, $\mathrm{P}=\mathrm{u}$ which gives $\mathcal{H}'_1(p^+_1)>0$ and $\mathcal{H}'_1(p^\mathrm{u}_1)<0$.
This occurs in the type V and VI processes.
In type V, the u mode is non-interacting and can be dropped from both sides of \eqref{energy_current4}.
Consequently, superradiance always occurs in type V scattering.
In type VI, the u mode is interacting and thus for amplification to occur at infinity, the total contribution of the $\mathcal{T}$ terms must be negative.
This is not always the case as we shall see in the next section.

Since the onset of type V scattering occurs below the lower critical frequency, the condition $\omega<\omega_\star$ is a necessary condition for superradiance in deep water.
The lower critical frequency can be brought to a form reminiscent of the shallow water condition \eqref{SR_cond1} by noting that the location of the inflection point is $r_\star=(6^\frac{1}{2}/2^\frac{1}{6})m^\frac{1}{3}$.
The condition becomes,
\begin{equation} \label{SR_cond3}
\omega < \omega_\star = \frac{m(C-2^{3/2})}{r_\star^2}.
\end{equation}
Whilst \eqref{SR_cond3} is a necessary condition for amplification, it is no longer sufficient to observe this amplification at infinity.
The reason for this is that in type VI, the total contribution of the $\mathcal{T}$ terms must be negative as previously mentioned.
Considering the type VI diagram in Fig.~\ref{fig:Feynmann_super}, this has a simple interpretation.
Even though amplification occurs at $r_3$ when the $-$ and $+$ modes scatter, the $+$ mode must scatter with the u mode before it appears at $r=R$.
If the distance between $r_4$ and $r_5$ is too large, then most of the amplified wave will be reflected back into the centre of the vortex.
Hence, to \textit{guarantee} amplification at infinity (within the WKB approximation) one must require,
\begin{equation} \label{SR_cond2}
\omega<\mathrm{min}(\omega_\star,\omega_\mathrm{LR}^+),
\end{equation}
which prevents the $+$ and u modes from scattering.
The region of parameter space in which this is satisfied is shaded in Fig~\ref{fig:ParamSpace_Types}.
Note that amplification at infinity can still occur just above the light ring frequency if the distance between $r_4$ and $r_5$ is sufficiently small.
However, we shall now see that the reflection coefficient in type VI decreases exponentially with the width of the tunnelling region.
Hence, \eqref{SR_cond2} should provide a useful working bound on when to expect superradiance in deep water systems.

\subsection{Reflection coefficients}

\begin{figure} 
\centering
\includegraphics[width=\linewidth]{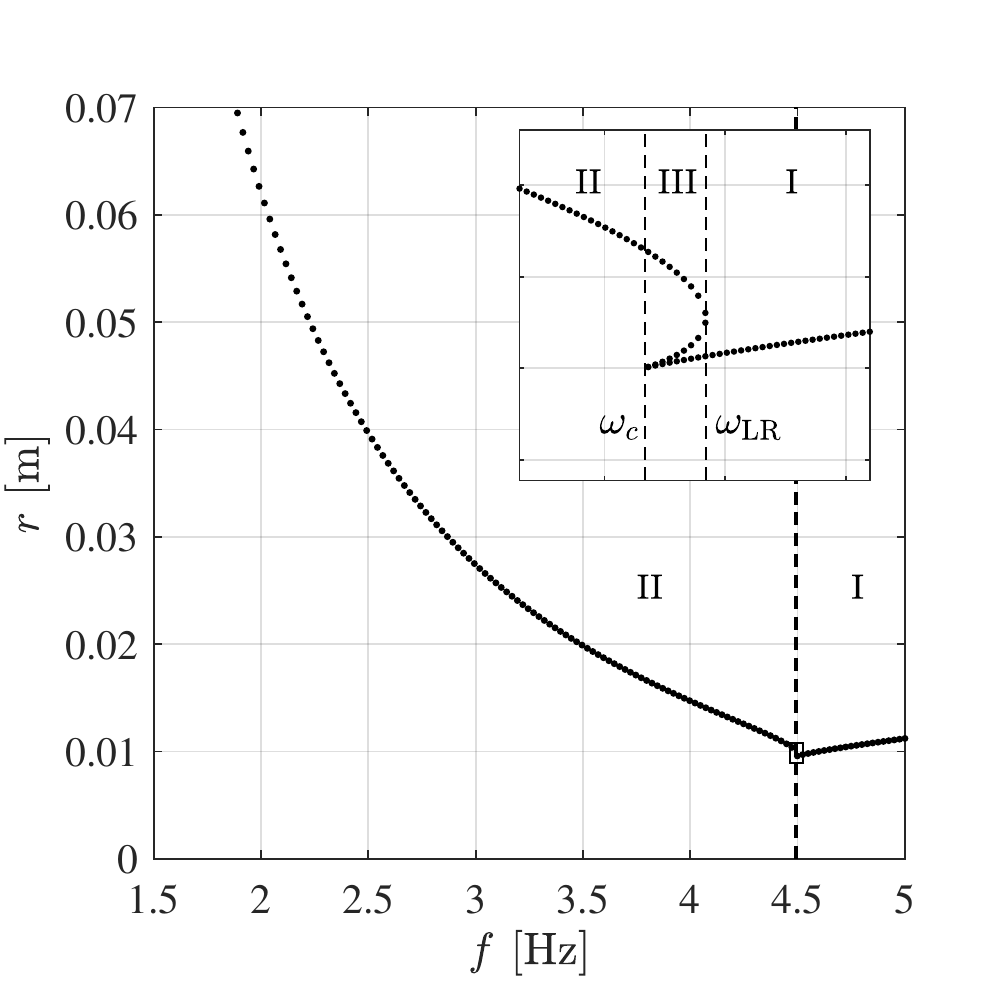}
\caption{The location of the turning points for the $m=1$ mode for a non-rotating flow with $D=9\times 10^{-4}~\mathrm{m/s}$. 
Axes are in dimensional units.
The small box close to 4.5 Hz is enlarged and displayed on the inset to resolve the narrow frequency range in which 3 turning points are present.
The type of scattering that occurs depends on the number of real turning points and is indicated on the figure.
Due to the symmetry in $m$ in non-rotating flows, the turning points for $m=-1$ are identical.
Furthermore, the turning points for higher $m$ follow the same trend as shown here, with the values of $r$ and $f$ scaled by a factor of $|m|^\frac{1}{3}$.
} \label{fig:TPs_equalC}
\end{figure}

\begin{figure} 
\centering
\includegraphics[width=\linewidth]{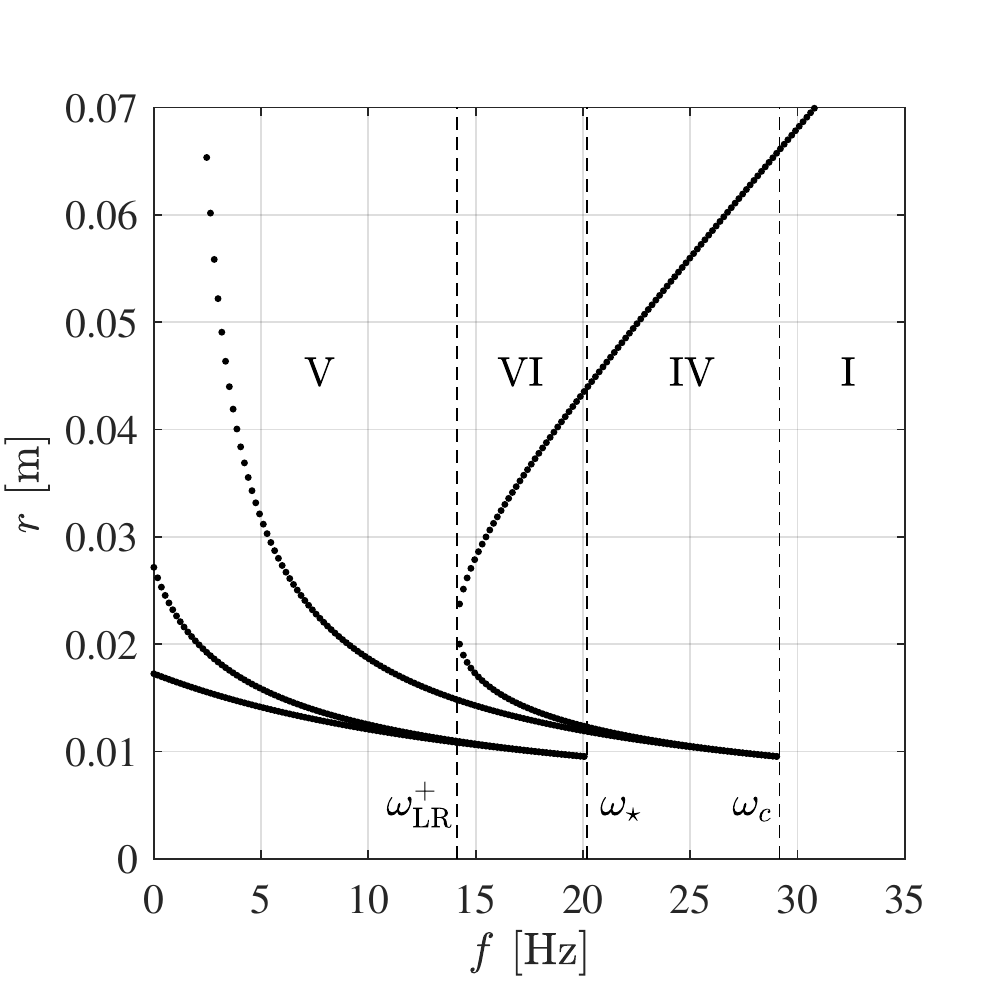}
\caption{Location of the real turning points as a function of frequency for the $m=1$ mode for the flow parameters used in the experiments of \cite{torres2017rotational}, i.e. $C=1.4\times 10^{-2}~\mathrm{m^2/s}$ and $D=9\times 10^{-4}~\mathrm{m^2/s}$.
} \label{fig:turning_points}
\end{figure}

The reflection coefficient \eqref{R_coeff} is computed using scattering matrix $\mathcal{M}$ which relates the mode amplitudes at $r_1$ to those at $R$,
\begin{equation} \label{scatterMat}
\begin{pmatrix}
A_1^\mathrm{u} \\ A_1^+ \\ A_1^- \\ A_1^\mathrm{d}
\end{pmatrix} = \mathcal{M} \begin{pmatrix}
A_R^\mathrm{u} \\ A_R^+ \\ A_R^- \\ A_R^\mathrm{d}
\end{pmatrix}.
\end{equation}
The detailed form of $\mathcal{M}$ will depend on the type of scattering taking place and is in general quite complicated.
An example calculation is given in Appendix B, for type V scattering.
The resulting expressions for $\mathcal{R}$ can be obtained by a much simpler method by inspecting the Feynman diagrams in Fig.~\ref{fig:Feynmann_super}.
The associated Feynman rules for the reflection coefficient are,
\begin{itemize}
\item Draw all possible paths which connect the $j^\mathrm{th}$ and $+$ mode at $R$ by following the arrows.
Each such path contributes a term to the reflection coefficient.
\item For each path, write down the ratio of the current of the $j^\mathrm{th}$ mode to that of the incident mode.
This is $1$ for the $-$ mode and $\mathcal{I}^\mathrm{u,d}$ for the u and d modes.
\item At each vertex, multiply by the local scattering coefficient given in \eqref{local_scatter}.
\item Multiply by the phase difference at the endpoints of the path given in \eqref{phase_diff}.
\item For the u and the d modes, multiply by a factor $(-1)$.
\end{itemize}
The resulting expressions for the different scattering processes are,
\begin{subequations}
\begin{align}
\mathrm{I}:~\mathcal{R} = \ & -e^{i(\varphi^\mathrm{u}-\pi/2)}\mathcal{I}^\mathrm{u}, \label{refl_type1} \\
\mathrm{II}:~\mathcal{R} = \ & e^{i(\varphi^--\pi/2)}, \label{refl_type2} \\ 
\mathrm{III}:~\mathcal{R} = \ & e^{i\varphi^-}\mathbf{R}^-_{23} - e^{i(\varphi^\mathrm{u}-\pi/2)}\mathbf{T}^-_{23}\mathcal{I}^\mathrm{u}, \label{refl_type3} \\
\mathrm{IV}:~\mathcal{R} = \ & e^{i(\varphi^--\pi/2)}\mathbf{T}^-_{23} - e^{i\varphi^\mathrm{u}}\mathbf{R}^-_{23}\mathcal{I}^\mathrm{u}, \label{refl_type4} \\
\mathrm{V}:~\mathcal{R} = \ & e^{i\varphi^-}\mathbf{R}^+_{23} - e^{i(\varphi^\mathrm{d}-\pi/2)}\mathbf{T}^+_{23}\mathcal{I}^\mathrm{d}, \label{refl_type5} \\
\mathrm{VI}:~\mathcal{R} = \ & e^{i\varphi^-}\mathbf{R}^+_{23}\mathbf{T}^-_{45} - e^{i\varphi^\mathrm{u}}\mathbf{R}^-_{45}\mathcal{I}^\mathrm{u} \label{refl_type6} \\
& \qquad \qquad - e^{i(\varphi^\mathrm{d}-\pi/2)}\mathbf{T}^+_{23}\mathbf{T}^-_{45}\mathcal{I}^\mathrm{d}. \nonumber
\end{align}
\end{subequations}
In these expressions, the local scattering coefficients are defined,
\begin{equation} \label{local_scatter}
\mathbf{R}^\pm_{ab} = e^{-\frac{i\pi}{2}}\left(\frac{1+\tfrac{1}{4}f_{ab}^2}{1-\tfrac{1}{4}f_{ab}^2}\right)^{\pm 1}, \; \mathbf{T}^\pm_{ab} = e^{-\frac{i}{4}(\pi\pm\pi)}\frac{f_{ab}}{1\mp\tfrac{1}{4}f_{ab}^2},
\end{equation} 
with $f_{ab}$ given in \eqref{LocalScatter2}.
The local reflection coefficient at $r_1$ is just $e^{-i\pi/2}$.
The phase factor $\varphi^j$ is the phase difference between the endpoints of the paths and is given by,
\begin{equation} \label{phase_diff}
\varphi = \mathrm{Re}\left[\int^R_{r_a}[p]^{\{\mathrm{path 2}\}}_{\{\mathrm{path 1}\}}dr\right],
\end{equation}
where path 1 starts at $R$ on the $j^\mathrm{th}$ mode and runs out to $r_a$, and path 2 runs from $r_a$ to the $+$ mode at $R$.
As an example, the contribution of the $-$ mode to type VI scattering has the following phase factor,
\begin{equation}
\int^R_{r_3}[p]_{\{-\}}^{\{+\uparrow +\}} = \int^R_{r_5}p^+dr + \int^{r_5}_{r_4}p^\uparrow dr + \int^{r_4}_{r_3}p^+dr - \int^R_{r_3}p^-dr
\end{equation}

As mentioned in the previous section, the d mode carries negative energy at large $r$ and is non-physical, thus, we can set $\mathcal{I}^\mathrm{d}=0$.
The u mode has positive energy and therefore does in principle contribute to $\mathcal{R}$.
However, in Appendix C, we argue that modes with large $k$ are heavily damped.
Since the growth of $k^\mathrm{u}$ is unbounded with increasing $r$, we can safely assume that any u modes sent in from afar will have dissipated away by the time they reach the vortex core.
Hence, we may also set $\mathcal{I}^\mathrm{u}=0$.
In this case, the formulae in \eqref{refl_type1} to \eqref{refl_type6} are telling us the following.
In type I scattering, no reflection occurs whereas in type II, one finds complete reflection.
Types III and V correspond to the same kind of scattering between $-$ and $+$ modes that occurs in shallow water, with type V being the superradiant case.
Types IV and VI include an additional interaction with the u channel.
In these cases, the in-going mode is reflected but must transmit back through an evanescent region before appearing at $r=R$.
As such, the reflection coefficient is exponentially suppressed for these cases.
In type VI, the $+$ mode gets superradiantly amplified at small $r$ but most of the extracted energy is reflected back into the vortex core by scattering with the u channel.

Finally, we note that the sharp transition between the use of different expressions in \eqref{refl_type1}-\eqref{refl_type6} is only an artefact of the approximation and, in reality, is smoothed over by backscattering off the inhomogeneous flow which couples the different modes even in the absence of turning points.
Such backscattering is of course exponentially suppressed when the difference between the $p^j$ is large \cite{coutant2014undulations}, however, it can become significant when two $p^j$ become close in the $(r,p)$ plane.
This occurs, in particular, in the vicinity of the light-ring frequencies shown in Fig.~\ref{fig:ParamSpace_Types}.
Near these curves, a saddle point approximation \cite{torres2020estimate} could be used to smooth over the discontinuities in $\mathcal{R}$.
This will be explored in future work.

\subsection{Non-rotating flow}

\begin{figure} 
\centering
\includegraphics[width=\linewidth]{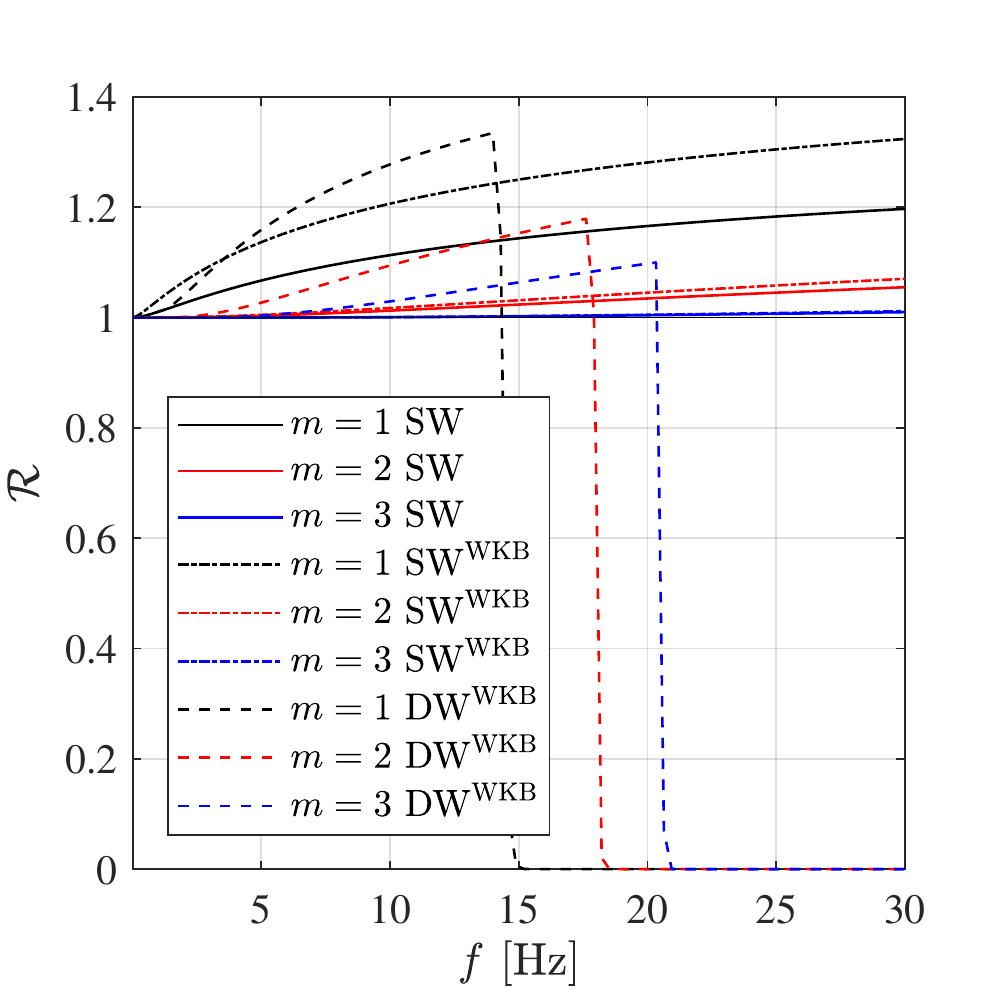}
\caption{Comparison between the reflection coefficient in shallow water (SW) and deep water (DW) for flow parameters of \cite{torres2017rotational}, i.e. $C=1.4\times 10^{-2}~\mathrm{m^2/s}$ and $D=9\times 10^{-4}~\mathrm{m^2/s}$.
The solid lines are the exact coefficients in shallow water computed from numerical simulation.
The broken lines are the SW WKB coefficients using \eqref{ReflWKB} and the dashed lines are the DW WKB coefficients using \eqref{refl_type1} to \eqref{refl_type6}.
Dispersion is shown to be able to increase the size of the reflection coefficient for large enough frequencies.
The cut-off frequency however is much lower than in shallow water, and agrees with the bound in \eqref{SR_cond2}.
} \label{fig:Compare}
\end{figure}

\begin{figure} 
\centering
\includegraphics[width=\linewidth]{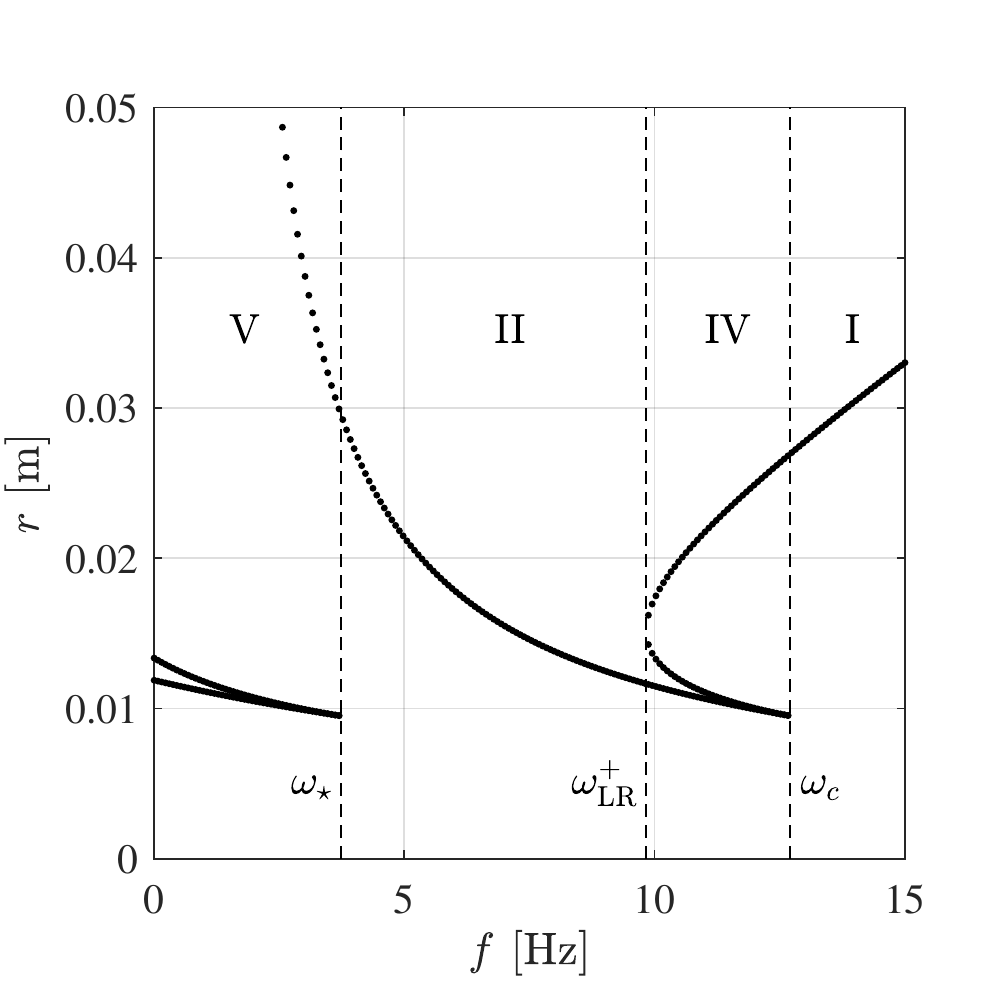}
\caption{Location of the real turning points as a function of frequency for the $m=1$ mode using the flow parameters $C=4.7\times 10^{-3}~\mathrm{m^2/s}$ and $D=9\times 10^{-4}~\mathrm{m^2/s}$.
} \label{fig:turning_points2}
\end{figure}

In the limit of vanishing rotation, the frequencies in \eqref{LightRings} and \eqref{CritFreqs} reduce to,
\begin{equation}
\begin{split}
\omega_c = & \ -\omega_\star = \frac{2^\frac{5}{6}}{3}|m|^\frac{1}{3} \approx 0.594~ |m|^\frac{1}{3}, \\
\omega_\mathrm{LR} = & \ \omega^\pm_\mathrm{LR} = \frac{4^\frac{1}{3}3}{8}|m|^\frac{1}{3} \approx 0.595~|m|^\frac{1}{3}.
\end{split}
\end{equation}
Notice, the light ring frequencies are the same for co- and counter-rotating modes, since in the absence of rotation the dispersion relation is invariant under $m\to-m$.
Since the lower critical frequency is negative, it plays no role in determining the scattering.

The location of the real turning points is displayed as a function of frequency in Fig.~\ref{fig:TPs_equalC}. 
Using this plot along with the \eqref{refl_type1} to \eqref{refl_type3}, one can predict the form of spectrum for the reflection coefficient.
Below $\omega_c$, $\mathcal{R}$ will be close to unity (type II) and then drop quickly toward zero between $\omega_c$ and $\omega_\mathrm{LR}$ (type III).
Above $\omega_\mathrm{LR}$, $\mathcal{R}$ will essentially be zero (type I).

\subsection{Rotating flow}

For rotating flows, the dependence of the turning points on frequency depends on the location in parameter space (see Fig.~\ref{fig:ParamSpace_Types}) which is dictated by the value of $C$.
Focussing on $m>0$, we give two examples for different values of $C$. 

For the first case, we take $C\approx 15.6$ in dimensionless variables.
This corresponds to the flow parameters used in the experiments of \cite{torres2017rotational}.
As the frequency is increased from zero, the system transitions through scattering types V, VI, IV and I.
Through this transition, the reflected mode is amplified (V), amplified but suppressed by further reflection (VI), not amplified and suppressed (IV) and finally negligible (I).
The real turning points for this case are shown in Fig.~\ref{fig:turning_points}.
The the reflection coefficient is displayed for the lowest three $m>0$ modes in Fig.~\ref{fig:Compare} and compared to the shallow water results.
These show that dispersion can result in more amplification, but the cut-off frequency where amplification ceases is lower than in shallow water.
The amount of amplification is below $20\%$ in the range 0 to 5 Hz, which is in the same ball-park as results in \cite{torres2017rotational}.

\begin{figure} 
\centering
\includegraphics[width=\linewidth]{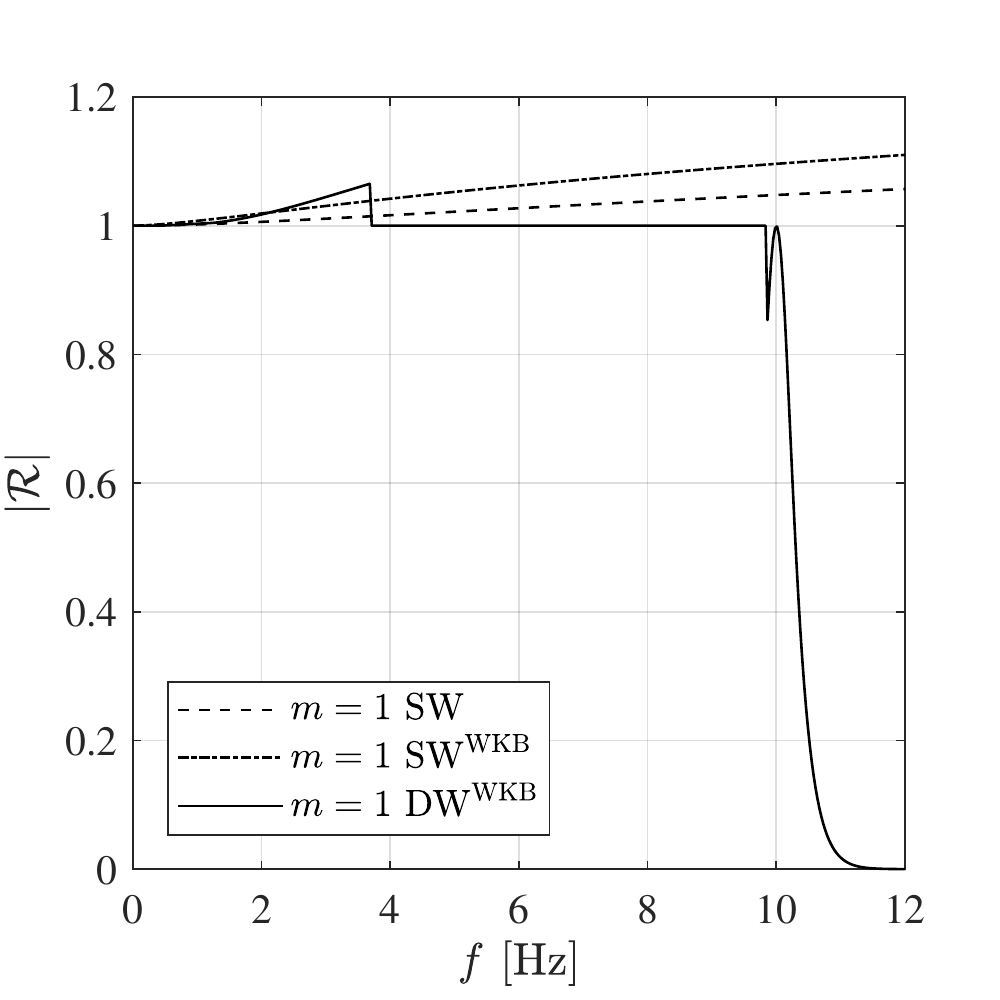}
\caption{The deep water reflection coefficients for the flow parameters $C=4.7\times 10^{-3}~\mathrm{m^2/s}$ and $D=9\times 10^{-4}~\mathrm{m^2/s}$, corresponding to a reduction of the rotation parameter in \cite{torres2017rotational} by a factor of 3.
Higher $m$-modes show qualitatively similar behaviour, albeit with less amplification.
The discontinuities at $\omega_\star\approx 3.7~\mathrm{Hz}$ and $\omega_\mathrm{LR}\approx 9.8~\mathrm{Hz}$ are artefacts of the approximation used here and in reality are smoothed over by backscattering.
The transition from amplification to pure reflection is a testable consequence of dispersion and should be observable within the frequency range probed by \cite{torres2017rotational}.
} \label{fig:LowRotation}
\end{figure}

The next case we consider corresponds to $C\approx 5.2$.
In this case, the system transitions through scattering types V, II, IV and I as the frequency is increased.
The initial and final behaviour of $\mathcal{R}$ is the same as the previous example, however, the intermediate type II region means there is a prolonged range in which $\mathcal{R}$ is approximately unity.
This contrasts the shallow water behaviour where the reflection coefficient always quick drops below 1 as the limiting frequency is surpassed.
The turning points for this scenario are plotted in Fig.~\ref{fig:turning_points2} and the reflection coefficients for the lowest lying $m>0$ modes are shown in Fig.~\ref{fig:LowRotation}.
For this set of flow parameters, the superradiant cut-off is within the frequency range probed by the experiment of \cite{torres2017rotational}.
Thus, by decreasing the rotation parameter in their experiment by a factor 3, it may be possible to test the predictions of Fig.~\ref{fig:LowRotation}.

\section{Conclusion}

In this work, we have developed a framework to analyse wave scattering in inhomogeneous systems.
The method involves treating the waves as semiclassical particles then tracing out the paths of these particles through the phase space.
Since particle-like and wave-like behaviour coincide for short wavelengths, these methods become increasingly accurate for high momentum modes in the system (in particular, high angular momentum).
We have then applied this framework to study the scattering of deep water gravity waves with a rotating, draining vortex flow.
This example was chosen due it's relevance for the experiments in \cite{torres2017rotational}.
However, the method can be applied to a wide variety of systems where the governing equation of motion is of the form \eqref{wave_equation}.
For example, Ho\v{r}ava gravity \cite{sotiriou2011lower,barausse2011black,barausse2013black} also exhibits a modified dispersion relation of this form.

The key finding of this study is that, in the deep water regime, the superradiance condition becomes that in \eqref{SR_cond2}.
We showed that the lower critical frequency $\omega_\star$ plays the role of the usual superradiance bound, determining when incident $\omega>0$ modes are amplified.
This, however, is not the full story, since the presence of extra modes in the system leads to other possible interactions.
Namely, amplified modes can be re-scattered by short-wavelength modes back into the vortex core, preventing them from extracting energy from the system.
This is a novel feature of dispersive systems that is completely absent in the shallow water approximation.
It so happens that the relevant frequency controlling this re-scattering is the well-known light-ring frequency \cite{torres2018waves}.
We expect this behaviour to not be limited to just the deep water regime, but rather a generic characteristic of sub-luminal dispersion relations.

Another novel feature of dispersive gravity waves in the DBT is that the propagation of long wavelength modes is prohibited in the vortex core in the frequency range $\omega_\star<\omega<\omega_c$.
This has testable consequences which should be observable within the frequency range probed by the experiments in \cite{torres2017rotational}.
In particular, Fig.~\ref{fig:LowRotation} demonstrates that one should observe complete reflection above the superradiant cut-off, if their circulation is slowed by a factor of 3.
Indeed, one of the motivations for this study was to explain the spectrum for the reflection coefficient in the experiments of \cite{torres2017rotational}.
Although we now have a framework to study scattering when the system is dispersive, we are still a few steps away from realising this goal.
In particular, dissipation, free surface gradients and vorticity have yet to be incorporated fully into the theoretical description.
In Appendix C, we have shown how dissipation can be described in 1D homogeneous fluid flows, and the effects of free surface gradients and vorticity have been studied in shallow water in \cite{richartz2015rotating} and \cite{patrick2018quasibound} respectively.
The inclusion of these effects into our formalism is a necessary step forward to make the connection to the on-going experimental efforts.

Finally, although in this work we have only included mode mixing near real turning points, our formalism can easily be extended to incorporate more sophisticated methods.
For example, mode mixing can be estimated in the vicinity of complex turning points \cite{coutant2016imprint}.
Furthermore, it is well-known that modes can mix around saddle points in phase space \cite{tracy2014ray}, and this method has recently been applied to estimate the reflection coefficients of the counter-rotating modes in the DBT \cite{torres2020estimate}.

\acknowledgements
SP acknowledges partial support provided by the Leverhulme Trust (Grant No. RPG-2016-233).
SW acknowledges financial support provided under the Paper Enhancement Grant at the University of Nottingham, the Royal Society University Research Fellow (UF120112), the Nottingham Advanced Research Fellow (A2RHS2), the Royal Society Enhancement Grant (RGF/EA/180286) and the EPSRC Project Grant (EP/P00637X/1). SW acknowledges partial support from STFC consolidated grant No. ST/P000703/.

\bibliographystyle{apsrev4-1}
\bibliography{superradiance_arXiv.bbl}

\begin{thebibliography}{66}%
\makeatletter
\providecommand \@ifxundefined [1]{%
 \@ifx{#1\undefined}
}%
\providecommand \@ifnum [1]{%
 \ifnum #1\expandafter \@firstoftwo
 \else \expandafter \@secondoftwo
 \fi
}%
\providecommand \@ifx [1]{%
 \ifx #1\expandafter \@firstoftwo
 \else \expandafter \@secondoftwo
 \fi
}%
\providecommand \natexlab [1]{#1}%
\providecommand \enquote  [1]{``#1''}%
\providecommand \bibnamefont  [1]{#1}%
\providecommand \bibfnamefont [1]{#1}%
\providecommand \citenamefont [1]{#1}%
\providecommand \href@noop [0]{\@secondoftwo}%
\providecommand \href [0]{\begingroup \@sanitize@url \@href}%
\providecommand \@href[1]{\@@startlink{#1}\@@href}%
\providecommand \@@href[1]{\endgroup#1\@@endlink}%
\providecommand \@sanitize@url [0]{\catcode `\\12\catcode `\$12\catcode
  `\&12\catcode `\#12\catcode `\^12\catcode `\_12\catcode `\%12\relax}%
\providecommand \@@startlink[1]{}%
\providecommand \@@endlink[0]{}%
\providecommand \url  [0]{\begingroup\@sanitize@url \@url }%
\providecommand \@url [1]{\endgroup\@href {#1}{\urlprefix }}%
\providecommand \urlprefix  [0]{URL }%
\providecommand \Eprint [0]{\href }%
\providecommand \doibase [0]{http://dx.doi.org/}%
\providecommand \selectlanguage [0]{\@gobble}%
\providecommand \bibinfo  [0]{\@secondoftwo}%
\providecommand \bibfield  [0]{\@secondoftwo}%
\providecommand \translation [1]{[#1]}%
\providecommand \BibitemOpen [0]{}%
\providecommand \bibitemStop [0]{}%
\providecommand \bibitemNoStop [0]{.\EOS\space}%
\providecommand \EOS [0]{\spacefactor3000\relax}%
\providecommand \BibitemShut  [1]{\csname bibitem#1\endcsname}%
\let\auto@bib@innerbib\@empty
\bibitem [{\citenamefont {Unruh}(1981)}]{unruh1981experimental}%
  \BibitemOpen
  \bibfield  {author} {\bibinfo {author} {\bibfnamefont {W.~G.}\ \bibnamefont
  {Unruh}},\ }\href@noop {} {\bibfield  {journal} {\bibinfo  {journal}
  {Physical Review Letters}\ }\textbf {\bibinfo {volume} {46}},\ \bibinfo
  {pages} {1351} (\bibinfo {year} {1981})}\BibitemShut {NoStop}%
\bibitem [{\citenamefont {Visser}(1993)}]{visser1993lorentzian}%
  \BibitemOpen
  \bibfield  {author} {\bibinfo {author} {\bibfnamefont {M.}~\bibnamefont
  {Visser}},\ }\href@noop {} {\bibfield  {journal} {\bibinfo  {journal} {arXiv:
  gr-qc/9311028}\ } (\bibinfo {year} {1993})},\ \Eprint
  {http://arxiv.org/abs/gr-qc/9311028} {arXiv:gr-qc/9311028 [gr-qc]}
  \BibitemShut {NoStop}%
\bibitem [{\citenamefont {Barcelo}\ \emph {et~al.}(2011)\citenamefont
  {Barcelo}, \citenamefont {Liberati},\ and\ \citenamefont
  {Visser}}]{barcelo2011analogue}%
  \BibitemOpen
  \bibfield  {author} {\bibinfo {author} {\bibfnamefont {C.}~\bibnamefont
  {Barcelo}}, \bibinfo {author} {\bibfnamefont {S.}~\bibnamefont {Liberati}}, \
  and\ \bibinfo {author} {\bibfnamefont {M.}~\bibnamefont {Visser}},\
  }\href@noop {} {\bibfield  {journal} {\bibinfo  {journal} {Living Reviews in
  Relativity}\ }\textbf {\bibinfo {volume} {14}},\ \bibinfo {pages} {3}
  (\bibinfo {year} {2011})}\BibitemShut {NoStop}%
\bibitem [{\citenamefont {Rousseaux}\ \emph {et~al.}(2008)\citenamefont
  {Rousseaux}, \citenamefont {Mathis}, \citenamefont {Ma{\"\i}ssa},
  \citenamefont {Philbin},\ and\ \citenamefont
  {Leonhardt}}]{rousseaux2008observation}%
  \BibitemOpen
  \bibfield  {author} {\bibinfo {author} {\bibfnamefont {G.}~\bibnamefont
  {Rousseaux}}, \bibinfo {author} {\bibfnamefont {C.}~\bibnamefont {Mathis}},
  \bibinfo {author} {\bibfnamefont {P.}~\bibnamefont {Ma{\"\i}ssa}}, \bibinfo
  {author} {\bibfnamefont {T.~G.}\ \bibnamefont {Philbin}}, \ and\ \bibinfo
  {author} {\bibfnamefont {U.}~\bibnamefont {Leonhardt}},\ }\href@noop {}
  {\bibfield  {journal} {\bibinfo  {journal} {New Journal of Physics}\ }\textbf
  {\bibinfo {volume} {10}},\ \bibinfo {pages} {053015} (\bibinfo {year}
  {2008})}\BibitemShut {NoStop}%
\bibitem [{\citenamefont {Weinfurtner}\ \emph {et~al.}(2011)\citenamefont
  {Weinfurtner}, \citenamefont {Tedford}, \citenamefont {Penrice},
  \citenamefont {Unruh},\ and\ \citenamefont
  {Lawrence}}]{weinfurtner2011measurement}%
  \BibitemOpen
  \bibfield  {author} {\bibinfo {author} {\bibfnamefont {S.}~\bibnamefont
  {Weinfurtner}}, \bibinfo {author} {\bibfnamefont {E.~W.}\ \bibnamefont
  {Tedford}}, \bibinfo {author} {\bibfnamefont {M.~C.~J.}\ \bibnamefont
  {Penrice}}, \bibinfo {author} {\bibfnamefont {W.~G.}\ \bibnamefont {Unruh}},
  \ and\ \bibinfo {author} {\bibfnamefont {G.~A.}\ \bibnamefont {Lawrence}},\
  }\href {\doibase 10.1103/PhysRevLett.106.021302} {\bibfield  {journal}
  {\bibinfo  {journal} {Physical Review Letters}\ }\textbf {\bibinfo {volume}
  {106}},\ \bibinfo {pages} {021302} (\bibinfo {year} {2011})}\BibitemShut
  {NoStop}%
\bibitem [{\citenamefont {Weinfurtner}\ \emph {et~al.}(2013)\citenamefont
  {Weinfurtner}, \citenamefont {Tedford}, \citenamefont {Penrice},
  \citenamefont {Unruh},\ and\ \citenamefont
  {Lawrence}}]{weinfurtner2013classical}%
  \BibitemOpen
  \bibfield  {author} {\bibinfo {author} {\bibfnamefont {S.}~\bibnamefont
  {Weinfurtner}}, \bibinfo {author} {\bibfnamefont {E.~W.}\ \bibnamefont
  {Tedford}}, \bibinfo {author} {\bibfnamefont {M.~C.~J.}\ \bibnamefont
  {Penrice}}, \bibinfo {author} {\bibfnamefont {W.~G.}\ \bibnamefont {Unruh}},
  \ and\ \bibinfo {author} {\bibfnamefont {G.~A.}\ \bibnamefont {Lawrence}},\
  }in\ \href@noop {} {\emph {\bibinfo {booktitle} {Analogue Gravity
  Phenomenology}}}\ (\bibinfo  {publisher} {Springer},\ \bibinfo {year}
  {2013})\ pp.\ \bibinfo {pages} {167--180}\BibitemShut {NoStop}%
\bibitem [{\citenamefont {Euv{\'e}}\ \emph {et~al.}(2016)\citenamefont
  {Euv{\'e}}, \citenamefont {Michel}, \citenamefont {Parentani}, \citenamefont
  {Philbin},\ and\ \citenamefont {Rousseaux}}]{euve2016observation}%
  \BibitemOpen
  \bibfield  {author} {\bibinfo {author} {\bibfnamefont {L.~P.}\ \bibnamefont
  {Euv{\'e}}}, \bibinfo {author} {\bibfnamefont {F.}~\bibnamefont {Michel}},
  \bibinfo {author} {\bibfnamefont {R.}~\bibnamefont {Parentani}}, \bibinfo
  {author} {\bibfnamefont {T.~G.}\ \bibnamefont {Philbin}}, \ and\ \bibinfo
  {author} {\bibfnamefont {G.}~\bibnamefont {Rousseaux}},\ }\href@noop {}
  {\bibfield  {journal} {\bibinfo  {journal} {Physical Review Letters}\
  }\textbf {\bibinfo {volume} {117}},\ \bibinfo {pages} {121301} (\bibinfo
  {year} {2016})}\BibitemShut {NoStop}%
\bibitem [{\citenamefont {Euv\'e}\ \emph {et~al.}(2020)\citenamefont {Euv\'e},
  \citenamefont {Robertson}, \citenamefont {James}, \citenamefont {Fabbri},\
  and\ \citenamefont {Rousseaux}}]{euve2020scattering}%
  \BibitemOpen
  \bibfield  {author} {\bibinfo {author} {\bibfnamefont {L.~P.}\ \bibnamefont
  {Euv\'e}}, \bibinfo {author} {\bibfnamefont {S.}~\bibnamefont {Robertson}},
  \bibinfo {author} {\bibfnamefont {N.}~\bibnamefont {James}}, \bibinfo
  {author} {\bibfnamefont {A.}~\bibnamefont {Fabbri}}, \ and\ \bibinfo {author}
  {\bibfnamefont {G.}~\bibnamefont {Rousseaux}},\ }\href {\doibase
  10.1103/PhysRevLett.124.141101} {\bibfield  {journal} {\bibinfo  {journal}
  {Physical Review Letters}\ }\textbf {\bibinfo {volume} {124}},\ \bibinfo
  {pages} {141101} (\bibinfo {year} {2020})}\BibitemShut {NoStop}%
\bibitem [{\citenamefont {Torres}\ \emph {et~al.}(2020)\citenamefont {Torres},
  \citenamefont {Patrick}, \citenamefont {Richartz},\ and\ \citenamefont
  {Weinfurtner}}]{torres2018application}%
  \BibitemOpen
  \bibfield  {author} {\bibinfo {author} {\bibfnamefont {T.}~\bibnamefont
  {Torres}}, \bibinfo {author} {\bibfnamefont {S.}~\bibnamefont {Patrick}},
  \bibinfo {author} {\bibfnamefont {M.}~\bibnamefont {Richartz}}, \ and\
  \bibinfo {author} {\bibfnamefont {S.}~\bibnamefont {Weinfurtner}},\ }\href
  {\doibase 10.1103/PhysRevLett.125.011301} {\bibfield  {journal} {\bibinfo
  {journal} {Physical Review Letters}\ }\textbf {\bibinfo {volume} {125}},\
  \bibinfo {pages} {011301} (\bibinfo {year} {2020})}\BibitemShut {NoStop}%
\bibitem [{\citenamefont {Sch{\"u}tzhold}\ and\ \citenamefont
  {Unruh}(2002)}]{schutzhold2002gravity}%
  \BibitemOpen
  \bibfield  {author} {\bibinfo {author} {\bibfnamefont {R.}~\bibnamefont
  {Sch{\"u}tzhold}}\ and\ \bibinfo {author} {\bibfnamefont {W.~G.}\
  \bibnamefont {Unruh}},\ }\href@noop {} {\bibfield  {journal} {\bibinfo
  {journal} {Physical Review D}\ }\textbf {\bibinfo {volume} {66}},\ \bibinfo
  {pages} {044019} (\bibinfo {year} {2002})}\BibitemShut {NoStop}%
\bibitem [{\citenamefont {Corley}\ and\ \citenamefont
  {Jacobson}(1996)}]{corley1996spectrum}%
  \BibitemOpen
  \bibfield  {author} {\bibinfo {author} {\bibfnamefont {S.}~\bibnamefont
  {Corley}}\ and\ \bibinfo {author} {\bibfnamefont {T.}~\bibnamefont
  {Jacobson}},\ }\href {\doibase 10.1103/PhysRevD.54.1568} {\bibfield
  {journal} {\bibinfo  {journal} {Physical Review D}\ }\textbf {\bibinfo
  {volume} {54}},\ \bibinfo {pages} {1568} (\bibinfo {year}
  {1996})}\BibitemShut {NoStop}%
\bibitem [{\citenamefont {Macher}\ and\ \citenamefont
  {Parentani}(2009)}]{macher2009black}%
  \BibitemOpen
  \bibfield  {author} {\bibinfo {author} {\bibfnamefont {J.}~\bibnamefont
  {Macher}}\ and\ \bibinfo {author} {\bibfnamefont {R.}~\bibnamefont
  {Parentani}},\ }\href {\doibase 10.1103/PhysRevD.79.124008} {\bibfield
  {journal} {\bibinfo  {journal} {Physical Review D}\ }\textbf {\bibinfo
  {volume} {79}},\ \bibinfo {pages} {124008} (\bibinfo {year}
  {2009})}\BibitemShut {NoStop}%
\bibitem [{\citenamefont {Finazzi}\ and\ \citenamefont
  {Parentani}(2012)}]{finazzi2012hawking}%
  \BibitemOpen
  \bibfield  {author} {\bibinfo {author} {\bibfnamefont {S.}~\bibnamefont
  {Finazzi}}\ and\ \bibinfo {author} {\bibfnamefont {R.}~\bibnamefont
  {Parentani}},\ }\href {\doibase 10.1103/PhysRevD.85.124027} {\bibfield
  {journal} {\bibinfo  {journal} {Physical Review D}\ }\textbf {\bibinfo
  {volume} {85}},\ \bibinfo {pages} {124027} (\bibinfo {year}
  {2012})}\BibitemShut {NoStop}%
\bibitem [{\citenamefont {Coutant}\ \emph {et~al.}(2012)\citenamefont
  {Coutant}, \citenamefont {Parentani},\ and\ \citenamefont
  {Finazzi}}]{coutant2012black}%
  \BibitemOpen
  \bibfield  {author} {\bibinfo {author} {\bibfnamefont {A.}~\bibnamefont
  {Coutant}}, \bibinfo {author} {\bibfnamefont {R.}~\bibnamefont {Parentani}},
  \ and\ \bibinfo {author} {\bibfnamefont {S.}~\bibnamefont {Finazzi}},\ }\href
  {\doibase 10.1103/PhysRevD.85.024021} {\bibfield  {journal} {\bibinfo
  {journal} {Physical Review D}\ }\textbf {\bibinfo {volume} {85}},\ \bibinfo
  {pages} {024021} (\bibinfo {year} {2012})}\BibitemShut {NoStop}%
\bibitem [{\citenamefont {Coutant}\ and\ \citenamefont
  {Parentani}(2014{\natexlab{a}})}]{coutant2014hawking}%
  \BibitemOpen
  \bibfield  {author} {\bibinfo {author} {\bibfnamefont {A.}~\bibnamefont
  {Coutant}}\ and\ \bibinfo {author} {\bibfnamefont {R.}~\bibnamefont
  {Parentani}},\ }\href {\doibase 10.1103/PhysRevD.90.121501} {\bibfield
  {journal} {\bibinfo  {journal} {Physical Review D}\ }\textbf {\bibinfo
  {volume} {90}},\ \bibinfo {pages} {121501} (\bibinfo {year}
  {2014}{\natexlab{a}})}\BibitemShut {NoStop}%
\bibitem [{\citenamefont {Robertson}\ \emph {et~al.}(2016)\citenamefont
  {Robertson}, \citenamefont {Michel},\ and\ \citenamefont
  {Parentani}}]{robertson2016scattering}%
  \BibitemOpen
  \bibfield  {author} {\bibinfo {author} {\bibfnamefont {S.}~\bibnamefont
  {Robertson}}, \bibinfo {author} {\bibfnamefont {F.}~\bibnamefont {Michel}}, \
  and\ \bibinfo {author} {\bibfnamefont {R.}~\bibnamefont {Parentani}},\ }\href
  {\doibase 10.1103/PhysRevD.93.124060} {\bibfield  {journal} {\bibinfo
  {journal} {Physical Review D}\ }\textbf {\bibinfo {volume} {93}},\ \bibinfo
  {pages} {124060} (\bibinfo {year} {2016})}\BibitemShut {NoStop}%
\bibitem [{\citenamefont {Coutant}\ and\ \citenamefont
  {Parentani}(2014{\natexlab{b}})}]{coutant2014undulations}%
  \BibitemOpen
  \bibfield  {author} {\bibinfo {author} {\bibfnamefont {A.}~\bibnamefont
  {Coutant}}\ and\ \bibinfo {author} {\bibfnamefont {R.}~\bibnamefont
  {Parentani}},\ }\href@noop {} {\bibfield  {journal} {\bibinfo  {journal}
  {Physics of Fluids}\ }\textbf {\bibinfo {volume} {26}},\ \bibinfo {pages}
  {044106} (\bibinfo {year} {2014}{\natexlab{b}})}\BibitemShut {NoStop}%
\bibitem [{\citenamefont {Coutant}\ and\ \citenamefont
  {Weinfurtner}(2016)}]{coutant2016imprint}%
  \BibitemOpen
  \bibfield  {author} {\bibinfo {author} {\bibfnamefont {A.}~\bibnamefont
  {Coutant}}\ and\ \bibinfo {author} {\bibfnamefont {S.}~\bibnamefont
  {Weinfurtner}},\ }\href@noop {} {\bibfield  {journal} {\bibinfo  {journal}
  {Physical Review D}\ }\textbf {\bibinfo {volume} {94}},\ \bibinfo {pages}
  {064026} (\bibinfo {year} {2016})}\BibitemShut {NoStop}%
\bibitem [{\citenamefont {{Dolan}}\ \emph {et~al.}(2011)\citenamefont
  {{Dolan}}, \citenamefont {{Oliveira}},\ and\ \citenamefont
  {{Crispino}}}]{dolan2011AB}%
  \BibitemOpen
  \bibfield  {author} {\bibinfo {author} {\bibfnamefont {S.~R.}\ \bibnamefont
  {{Dolan}}}, \bibinfo {author} {\bibfnamefont {E.~S.}\ \bibnamefont
  {{Oliveira}}}, \ and\ \bibinfo {author} {\bibfnamefont {L.~C.~B.}\
  \bibnamefont {{Crispino}}},\ }\href {\doibase 10.1016/j.physletb.2011.06.013}
  {\bibfield  {journal} {\bibinfo  {journal} {Physics Letters B}\ }\textbf
  {\bibinfo {volume} {701}},\ \bibinfo {pages} {485} (\bibinfo {year}
  {2011})}\BibitemShut {NoStop}%
\bibitem [{\citenamefont {Dolan}\ \emph {et~al.}(2012)\citenamefont {Dolan},
  \citenamefont {Oliveira},\ and\ \citenamefont
  {Crispino}}]{dolan2012resonances}%
  \BibitemOpen
  \bibfield  {author} {\bibinfo {author} {\bibfnamefont {S.~R.}\ \bibnamefont
  {Dolan}}, \bibinfo {author} {\bibfnamefont {L.~A.}\ \bibnamefont {Oliveira}},
  \ and\ \bibinfo {author} {\bibfnamefont {L.~C.~B.}\ \bibnamefont
  {Crispino}},\ }\href@noop {} {\bibfield  {journal} {\bibinfo  {journal}
  {Physical Review D}\ }\textbf {\bibinfo {volume} {85}},\ \bibinfo {pages}
  {044031} (\bibinfo {year} {2012})}\BibitemShut {NoStop}%
\bibitem [{\citenamefont {Dolan}\ and\ \citenamefont
  {Oliveira}(2013)}]{dolan2013scattering}%
  \BibitemOpen
  \bibfield  {author} {\bibinfo {author} {\bibfnamefont {S.~R.}\ \bibnamefont
  {Dolan}}\ and\ \bibinfo {author} {\bibfnamefont {E.~S.}\ \bibnamefont
  {Oliveira}},\ }\href {\doibase 10.1103/PhysRevD.87.124038} {\bibfield
  {journal} {\bibinfo  {journal} {Physical Review D}\ }\textbf {\bibinfo
  {volume} {87}},\ \bibinfo {pages} {124038} (\bibinfo {year}
  {2013})}\BibitemShut {NoStop}%
\bibitem [{\citenamefont {Berti}\ \emph {et~al.}(2004)\citenamefont {Berti},
  \citenamefont {Cardoso},\ and\ \citenamefont {Lemos}}]{berti2004qnm}%
  \BibitemOpen
  \bibfield  {author} {\bibinfo {author} {\bibfnamefont {E.}~\bibnamefont
  {Berti}}, \bibinfo {author} {\bibfnamefont {V.}~\bibnamefont {Cardoso}}, \
  and\ \bibinfo {author} {\bibfnamefont {J.~P.~S.}\ \bibnamefont {Lemos}},\
  }\href@noop {} {\bibfield  {journal} {\bibinfo  {journal} {Physical Review
  D}\ }\textbf {\bibinfo {volume} {70}},\ \bibinfo {pages} {124006} (\bibinfo
  {year} {2004})}\BibitemShut {NoStop}%
\bibitem [{\citenamefont {Cardoso}\ \emph {et~al.}(2004)\citenamefont
  {Cardoso}, \citenamefont {Lemos},\ and\ \citenamefont
  {Yoshida}}]{cardoso2004qnm}%
  \BibitemOpen
  \bibfield  {author} {\bibinfo {author} {\bibfnamefont {V.}~\bibnamefont
  {Cardoso}}, \bibinfo {author} {\bibfnamefont {J.~P.~S.}\ \bibnamefont
  {Lemos}}, \ and\ \bibinfo {author} {\bibfnamefont {S.}~\bibnamefont
  {Yoshida}},\ }\href@noop {} {\bibfield  {journal} {\bibinfo  {journal}
  {Physical Review D}\ }\textbf {\bibinfo {volume} {70}},\ \bibinfo {pages}
  {124032} (\bibinfo {year} {2004})}\BibitemShut {NoStop}%
\bibitem [{\citenamefont {Torres}\ \emph {et~al.}(2019)\citenamefont {Torres},
  \citenamefont {Patrick}, \citenamefont {Richartz},\ and\ \citenamefont
  {Weinfurtner}}]{torres2019analogue}%
  \BibitemOpen
  \bibfield  {author} {\bibinfo {author} {\bibfnamefont {T.}~\bibnamefont
  {Torres}}, \bibinfo {author} {\bibfnamefont {S.}~\bibnamefont {Patrick}},
  \bibinfo {author} {\bibfnamefont {M.}~\bibnamefont {Richartz}}, \ and\
  \bibinfo {author} {\bibfnamefont {S.}~\bibnamefont {Weinfurtner}},\ }\href
  {\doibase 10.1088/1361-6382/ab3d48} {\bibfield  {journal} {\bibinfo
  {journal} {Classical and Quantum Gravity}\ }\textbf {\bibinfo {volume}
  {36}},\ \bibinfo {pages} {194002} (\bibinfo {year} {2019})}\BibitemShut
  {NoStop}%
\bibitem [{\citenamefont {{Basak}}\ and\ \citenamefont
  {{Majumdar}}(2003{\natexlab{a}})}]{basak2003superresonance}%
  \BibitemOpen
  \bibfield  {author} {\bibinfo {author} {\bibfnamefont {S.}~\bibnamefont
  {{Basak}}}\ and\ \bibinfo {author} {\bibfnamefont {P.}~\bibnamefont
  {{Majumdar}}},\ }\href@noop {} {\bibfield  {journal} {\bibinfo  {journal}
  {Classical and Quantum Gravity}\ }\textbf {\bibinfo {volume} {20}},\ \bibinfo
  {pages} {3907} (\bibinfo {year} {2003}{\natexlab{a}})}\BibitemShut {NoStop}%
\bibitem [{\citenamefont {{Basak}}\ and\ \citenamefont
  {{Majumdar}}(2003{\natexlab{b}})}]{basak2003reflection}%
  \BibitemOpen
  \bibfield  {author} {\bibinfo {author} {\bibfnamefont {S.}~\bibnamefont
  {{Basak}}}\ and\ \bibinfo {author} {\bibfnamefont {P.}~\bibnamefont
  {{Majumdar}}},\ }\href {\doibase 10.1088/0264-9381/20/13/335} {\bibfield
  {journal} {\bibinfo  {journal} {Classical and Quantum Gravity}\ }\textbf
  {\bibinfo {volume} {20}},\ \bibinfo {pages} {2929} (\bibinfo {year}
  {2003}{\natexlab{b}})}\BibitemShut {NoStop}%
\bibitem [{\citenamefont {Richartz}\ \emph {et~al.}(2015)\citenamefont
  {Richartz}, \citenamefont {Prain}, \citenamefont {Liberati},\ and\
  \citenamefont {Weinfurtner}}]{richartz2015rotating}%
  \BibitemOpen
  \bibfield  {author} {\bibinfo {author} {\bibfnamefont {M.}~\bibnamefont
  {Richartz}}, \bibinfo {author} {\bibfnamefont {A.}~\bibnamefont {Prain}},
  \bibinfo {author} {\bibfnamefont {S.}~\bibnamefont {Liberati}}, \ and\
  \bibinfo {author} {\bibfnamefont {S.}~\bibnamefont {Weinfurtner}},\
  }\href@noop {} {\bibfield  {journal} {\bibinfo  {journal} {Physical Review
  D}\ }\textbf {\bibinfo {volume} {91}},\ \bibinfo {pages} {124018} (\bibinfo
  {year} {2015})}\BibitemShut {NoStop}%
\bibitem [{\citenamefont {{Bekenstein}}\ and\ \citenamefont
  {{Schiffer}}(1998)}]{bekenstein1998many}%
  \BibitemOpen
  \bibfield  {author} {\bibinfo {author} {\bibfnamefont {J.~D.}\ \bibnamefont
  {{Bekenstein}}}\ and\ \bibinfo {author} {\bibfnamefont {M.}~\bibnamefont
  {{Schiffer}}},\ }\href@noop {} {\bibfield  {journal} {\bibinfo  {journal}
  {Physical Review D}\ }\textbf {\bibinfo {volume} {58}},\ \bibinfo {pages}
  {064014} (\bibinfo {year} {1998})}\BibitemShut {NoStop}%
\bibitem [{\citenamefont {{Brito}}\ \emph {et~al.}(2015)\citenamefont
  {{Brito}}, \citenamefont {{Cardoso}},\ and\ \citenamefont
  {{Pani}}}]{brito2015superradiance}%
  \BibitemOpen
  \bibfield  {author} {\bibinfo {author} {\bibfnamefont {R.}~\bibnamefont
  {{Brito}}}, \bibinfo {author} {\bibfnamefont {V.}~\bibnamefont {{Cardoso}}},
  \ and\ \bibinfo {author} {\bibfnamefont {P.}~\bibnamefont {{Pani}}},\
  }\href@noop {} {\bibfield  {journal} {\bibinfo  {journal} {Lecture Notes in
  Physics}\ }\textbf {\bibinfo {volume} {906}},\ \bibinfo {pages} {18}
  (\bibinfo {year} {2015})}\BibitemShut {NoStop}%
\bibitem [{\citenamefont {Ginzburg}\ and\ \citenamefont
  {Frank}(1947)}]{ginzburg1947doppler}%
  \BibitemOpen
  \bibfield  {author} {\bibinfo {author} {\bibfnamefont {V.~L.}\ \bibnamefont
  {Ginzburg}}\ and\ \bibinfo {author} {\bibfnamefont {I.~M.}\ \bibnamefont
  {Frank}},\ }\bibfield  {booktitle} {\emph {\bibinfo {booktitle} {Doklady
  Akademii Nauk SSSR}},\ }\href@noop {} {\ \textbf {\bibinfo {volume} {56}},\
  \bibinfo {pages} {583} (\bibinfo {year} {1947})}\BibitemShut {NoStop}%
\bibitem [{\citenamefont {Ginzburg}(1993)}]{ginzburg1993v}%
  \BibitemOpen
  \bibfield  {author} {\bibinfo {author} {\bibfnamefont {V.~L.}\ \bibnamefont
  {Ginzburg}},\ }\bibfield  {booktitle} {\emph {\bibinfo {booktitle} {Progress
  in optics}},\ }\href@noop {} {\ \textbf {\bibinfo {volume} {32}},\ \bibinfo
  {pages} {267} (\bibinfo {year} {1993})}\BibitemShut {NoStop}%
\bibitem [{\citenamefont {Dicke}(1954)}]{dicke1954coherence}%
  \BibitemOpen
  \bibfield  {author} {\bibinfo {author} {\bibfnamefont {R.~H.}\ \bibnamefont
  {Dicke}},\ }\href@noop {} {\bibfield  {journal} {\bibinfo  {journal}
  {Physical Review}\ }\textbf {\bibinfo {volume} {93}},\ \bibinfo {pages} {99}
  (\bibinfo {year} {1954})}\BibitemShut {NoStop}%
\bibitem [{\citenamefont {{Zel'Dovich}}(1971)}]{zeldovich1971generation}%
  \BibitemOpen
  \bibfield  {author} {\bibinfo {author} {\bibfnamefont {Y.~B.}\ \bibnamefont
  {{Zel'Dovich}}},\ }\href@noop {} {\bibfield  {journal} {\bibinfo  {journal}
  {ZhETF Pisma Redaktsiiu}\ }\textbf {\bibinfo {volume} {14}},\ \bibinfo
  {pages} {270} (\bibinfo {year} {1971})}\BibitemShut {NoStop}%
\bibitem [{\citenamefont {{Zel'Dovich}}(1972)}]{zeldovich1972amplification}%
  \BibitemOpen
  \bibfield  {author} {\bibinfo {author} {\bibfnamefont {Y.~B.}\ \bibnamefont
  {{Zel'Dovich}}},\ }\href@noop {} {\bibfield  {journal} {\bibinfo  {journal}
  {Soviet Journal of Experimental and Theoretical Physics}\ }\textbf {\bibinfo
  {volume} {35}},\ \bibinfo {pages} {1085} (\bibinfo {year}
  {1972})}\BibitemShut {NoStop}%
\bibitem [{\citenamefont {{McKenzie}}(1972)}]{mckenzie1972reflection}%
  \BibitemOpen
  \bibfield  {author} {\bibinfo {author} {\bibfnamefont {J.~F.}\ \bibnamefont
  {{McKenzie}}},\ }\href@noop {} {\bibfield  {journal} {\bibinfo  {journal}
  {Journal of Geophysical Research}\ }\textbf {\bibinfo {volume} {77}},\
  \bibinfo {pages} {2915} (\bibinfo {year} {1972})}\BibitemShut {NoStop}%
\bibitem [{\citenamefont {{Acheson}}(1976)}]{acheson1976overreflexion}%
  \BibitemOpen
  \bibfield  {author} {\bibinfo {author} {\bibfnamefont {D.~J.}\ \bibnamefont
  {{Acheson}}},\ }\href {\doibase 10.1017/S0022112076002206} {\bibfield
  {journal} {\bibinfo  {journal} {Journal of Fluid Mechanics}\ }\textbf
  {\bibinfo {volume} {77}},\ \bibinfo {pages} {433} (\bibinfo {year}
  {1976})}\BibitemShut {NoStop}%
\bibitem [{\citenamefont {{Kelley}}\ \emph {et~al.}(2007)\citenamefont
  {{Kelley}}, \citenamefont {{Triana}}, \citenamefont {{Zimmerman}},
  \citenamefont {{Tilgner}},\ and\ \citenamefont
  {{Lathrop}}}]{kelley2007inertial}%
  \BibitemOpen
  \bibfield  {author} {\bibinfo {author} {\bibfnamefont {D.~H.}\ \bibnamefont
  {{Kelley}}}, \bibinfo {author} {\bibfnamefont {S.~A.}\ \bibnamefont
  {{Triana}}}, \bibinfo {author} {\bibfnamefont {D.~S.}\ \bibnamefont
  {{Zimmerman}}}, \bibinfo {author} {\bibfnamefont {A.}~\bibnamefont
  {{Tilgner}}}, \ and\ \bibinfo {author} {\bibfnamefont {D.~P.}\ \bibnamefont
  {{Lathrop}}},\ }\href@noop {} {\bibfield  {journal} {\bibinfo  {journal}
  {Geophysical and Astrophysical Fluid Dynamics}\ }\textbf {\bibinfo {volume}
  {101}},\ \bibinfo {pages} {469} (\bibinfo {year} {2007})}\BibitemShut
  {NoStop}%
\bibitem [{\citenamefont {{Fridman}}\ \emph {et~al.}(2008)\citenamefont
  {{Fridman}}, \citenamefont {{Snezhkin}}, \citenamefont {{Chernikov}},
  \citenamefont {{Rylov}}, \citenamefont {{Titishov}},\ and\ \citenamefont
  {{Torgashin}}}]{fridman2008overreflection}%
  \BibitemOpen
  \bibfield  {author} {\bibinfo {author} {\bibfnamefont {A.~M.}\ \bibnamefont
  {{Fridman}}}, \bibinfo {author} {\bibfnamefont {E.~N.}\ \bibnamefont
  {{Snezhkin}}}, \bibinfo {author} {\bibfnamefont {G.~P.}\ \bibnamefont
  {{Chernikov}}}, \bibinfo {author} {\bibfnamefont {A.~Y.}\ \bibnamefont
  {{Rylov}}}, \bibinfo {author} {\bibfnamefont {K.~B.}\ \bibnamefont
  {{Titishov}}}, \ and\ \bibinfo {author} {\bibfnamefont {Y.~M.}\ \bibnamefont
  {{Torgashin}}},\ }\href@noop {} {\bibfield  {journal} {\bibinfo  {journal}
  {Physics Letters A}\ }\textbf {\bibinfo {volume} {372}},\ \bibinfo {pages}
  {4822} (\bibinfo {year} {2008})}\BibitemShut {NoStop}%
\bibitem [{\citenamefont {{Penrose}}\ and\ \citenamefont
  {{Floyd}}(1971)}]{penrose1971extraction}%
  \BibitemOpen
  \bibfield  {author} {\bibinfo {author} {\bibfnamefont {R.}~\bibnamefont
  {{Penrose}}}\ and\ \bibinfo {author} {\bibfnamefont {R.~M.}\ \bibnamefont
  {{Floyd}}},\ }\href@noop {} {\bibfield  {journal} {\bibinfo  {journal}
  {Nature Physical Science}\ }\textbf {\bibinfo {volume} {229}},\ \bibinfo
  {pages} {177} (\bibinfo {year} {1971})}\BibitemShut {NoStop}%
\bibitem [{\citenamefont {{Misner}}(1972)}]{misner1972stability}%
  \BibitemOpen
  \bibfield  {author} {\bibinfo {author} {\bibfnamefont {C.}~\bibnamefont
  {{Misner}}},\ }\href@noop {} {\bibfield  {journal} {\bibinfo  {journal}
  {Bulletin of the American Physical Society}\ }\textbf {\bibinfo {volume}
  {17}},\ \bibinfo {pages} {472} (\bibinfo {year} {1972})}\BibitemShut
  {NoStop}%
\bibitem [{\citenamefont {{Starobinski{\v i}}}(1973)}]{starobinsky1974waves}%
  \BibitemOpen
  \bibfield  {author} {\bibinfo {author} {\bibfnamefont {A.~A.}\ \bibnamefont
  {{Starobinski{\v i}}}},\ }\href@noop {} {\bibfield  {journal} {\bibinfo
  {journal} {Soviet Journal of Experimental and Theoretical Physics}\ }\textbf
  {\bibinfo {volume} {37}},\ \bibinfo {pages} {28} (\bibinfo {year}
  {1973})}\BibitemShut {NoStop}%
\bibitem [{\citenamefont {{Starobinski{\v i}}}\ and\ \citenamefont
  {{Churilov}}(1974)}]{starobinsky1974electro}%
  \BibitemOpen
  \bibfield  {author} {\bibinfo {author} {\bibfnamefont {A.~A.}\ \bibnamefont
  {{Starobinski{\v i}}}}\ and\ \bibinfo {author} {\bibfnamefont {S.~M.}\
  \bibnamefont {{Churilov}}},\ }\href@noop {} {\bibfield  {journal} {\bibinfo
  {journal} {Soviet Journal of Experimental and Theoretical Physics}\ }\textbf
  {\bibinfo {volume} {38}},\ \bibinfo {pages} {1} (\bibinfo {year}
  {1974})}\BibitemShut {NoStop}%
\bibitem [{\citenamefont {Hawking}(1974)}]{hawking1974explosions}%
  \BibitemOpen
  \bibfield  {author} {\bibinfo {author} {\bibfnamefont {S.~W.}\ \bibnamefont
  {Hawking}},\ }\href@noop {} {\bibfield  {journal} {\bibinfo  {journal}
  {Nature}\ }\textbf {\bibinfo {volume} {248}},\ \bibinfo {pages} {30}
  (\bibinfo {year} {1974})}\BibitemShut {NoStop}%
\bibitem [{\citenamefont {Bekenstein}(1994)}]{bekenstein1994entropy}%
  \BibitemOpen
  \bibfield  {author} {\bibinfo {author} {\bibfnamefont {J.~D.}\ \bibnamefont
  {Bekenstein}},\ }\href@noop {} {\bibfield  {journal} {\bibinfo  {journal}
  {Physical Review D}\ }\textbf {\bibinfo {volume} {49}},\ \bibinfo {pages}
  {1912} (\bibinfo {year} {1994})}\BibitemShut {NoStop}%
\bibitem [{\citenamefont {Brito}\ \emph {et~al.}(2017)\citenamefont {Brito},
  \citenamefont {Ghosh}, \citenamefont {Barausse}, \citenamefont {Berti},
  \citenamefont {Cardoso}, \citenamefont {Dvorkin}, \citenamefont {Klein},\
  and\ \citenamefont {Pani}}]{brito2017gravitational}%
  \BibitemOpen
  \bibfield  {author} {\bibinfo {author} {\bibfnamefont {R.}~\bibnamefont
  {Brito}}, \bibinfo {author} {\bibfnamefont {S.}~\bibnamefont {Ghosh}},
  \bibinfo {author} {\bibfnamefont {E.}~\bibnamefont {Barausse}}, \bibinfo
  {author} {\bibfnamefont {E.}~\bibnamefont {Berti}}, \bibinfo {author}
  {\bibfnamefont {V.}~\bibnamefont {Cardoso}}, \bibinfo {author} {\bibfnamefont
  {I.}~\bibnamefont {Dvorkin}}, \bibinfo {author} {\bibfnamefont
  {A.}~\bibnamefont {Klein}}, \ and\ \bibinfo {author} {\bibfnamefont
  {P.}~\bibnamefont {Pani}},\ }\href {\doibase 10.1103/PhysRevD.96.064050}
  {\bibfield  {journal} {\bibinfo  {journal} {Physical Review D}\ }\textbf
  {\bibinfo {volume} {96}},\ \bibinfo {pages} {064050} (\bibinfo {year}
  {2017})}\BibitemShut {NoStop}%
\bibitem [{\citenamefont {Baumann}\ \emph {et~al.}(2019)\citenamefont
  {Baumann}, \citenamefont {Chia},\ and\ \citenamefont
  {Porto}}]{baumann2019probing}%
  \BibitemOpen
  \bibfield  {author} {\bibinfo {author} {\bibfnamefont {D.}~\bibnamefont
  {Baumann}}, \bibinfo {author} {\bibfnamefont {H.~S.}\ \bibnamefont {Chia}}, \
  and\ \bibinfo {author} {\bibfnamefont {R.~A.}\ \bibnamefont {Porto}},\ }\href
  {\doibase 10.1103/PhysRevD.99.044001} {\bibfield  {journal} {\bibinfo
  {journal} {Physical Review D}\ }\textbf {\bibinfo {volume} {99}},\ \bibinfo
  {pages} {044001} (\bibinfo {year} {2019})}\BibitemShut {NoStop}%
\bibitem [{\citenamefont {Siemonsen}\ and\ \citenamefont
  {East}(2020)}]{siemonsen2020gravitational}%
  \BibitemOpen
  \bibfield  {author} {\bibinfo {author} {\bibfnamefont {N.}~\bibnamefont
  {Siemonsen}}\ and\ \bibinfo {author} {\bibfnamefont {W.~E.}\ \bibnamefont
  {East}},\ }\href {\doibase 10.1103/PhysRevD.101.024019} {\bibfield  {journal}
  {\bibinfo  {journal} {Physical Review D}\ }\textbf {\bibinfo {volume}
  {101}},\ \bibinfo {pages} {024019} (\bibinfo {year} {2020})}\BibitemShut
  {NoStop}%
\bibitem [{\citenamefont {Torres}\ \emph {et~al.}(2017)\citenamefont {Torres},
  \citenamefont {Patrick}, \citenamefont {Coutant}, \citenamefont {Richartz},
  \citenamefont {Tedford},\ and\ \citenamefont
  {Weinfurtner}}]{torres2017rotational}%
  \BibitemOpen
  \bibfield  {author} {\bibinfo {author} {\bibfnamefont {T.}~\bibnamefont
  {Torres}}, \bibinfo {author} {\bibfnamefont {S.}~\bibnamefont {Patrick}},
  \bibinfo {author} {\bibfnamefont {A.}~\bibnamefont {Coutant}}, \bibinfo
  {author} {\bibfnamefont {M.}~\bibnamefont {Richartz}}, \bibinfo {author}
  {\bibfnamefont {E.~W.}\ \bibnamefont {Tedford}}, \ and\ \bibinfo {author}
  {\bibfnamefont {S.}~\bibnamefont {Weinfurtner}},\ }\href@noop {} {\bibfield
  {journal} {\bibinfo  {journal} {Nature Physics}\ }\textbf {\bibinfo {volume}
  {13}},\ \bibinfo {pages} {833} (\bibinfo {year} {2017})}\BibitemShut
  {NoStop}%
\bibitem [{\citenamefont {B{\"u}hler}(2014)}]{buhler2014waves}%
  \BibitemOpen
  \bibfield  {author} {\bibinfo {author} {\bibfnamefont {O.}~\bibnamefont
  {B{\"u}hler}},\ }\href@noop {} {\emph {\bibinfo {title} {Waves and mean
  flows}}}\ (\bibinfo  {publisher} {Cambridge University Press},\ \bibinfo
  {year} {2014})\BibitemShut {NoStop}%
\bibitem [{\citenamefont {Berry}\ and\ \citenamefont
  {Mount}(1972)}]{berry1972semiclassical}%
  \BibitemOpen
  \bibfield  {author} {\bibinfo {author} {\bibfnamefont {M.~V.}\ \bibnamefont
  {Berry}}\ and\ \bibinfo {author} {\bibfnamefont {K.~E.}\ \bibnamefont
  {Mount}},\ }\href@noop {} {\bibfield  {journal} {\bibinfo  {journal} {Reports
  on Progress in Physics}\ }\textbf {\bibinfo {volume} {35}},\ \bibinfo {pages}
  {315} (\bibinfo {year} {1972})}\BibitemShut {NoStop}%
\bibitem [{\citenamefont {Tracy}\ \emph {et~al.}(2014)\citenamefont {Tracy},
  \citenamefont {Brizard}, \citenamefont {Richardson},\ and\ \citenamefont
  {Kaufman}}]{tracy2014ray}%
  \BibitemOpen
  \bibfield  {author} {\bibinfo {author} {\bibfnamefont {E.~R.}\ \bibnamefont
  {Tracy}}, \bibinfo {author} {\bibfnamefont {A.~J.}\ \bibnamefont {Brizard}},
  \bibinfo {author} {\bibfnamefont {A.~S.}\ \bibnamefont {Richardson}}, \ and\
  \bibinfo {author} {\bibfnamefont {A.~N.}\ \bibnamefont {Kaufman}},\
  }\href@noop {} {\emph {\bibinfo {title} {Ray Tracing and Beyond: Phase Space
  Methods in Plasma Wave Theory}}}\ (\bibinfo  {publisher} {Cambridge
  University Press},\ \bibinfo {year} {2014})\BibitemShut {NoStop}%
\bibitem [{\citenamefont {Arnowitt}\ \emph {et~al.}(1962)\citenamefont
  {Arnowitt}, \citenamefont {Deser},\ and\ \citenamefont
  {Misner}}]{arnowitt1962dynamics}%
  \BibitemOpen
  \bibfield  {author} {\bibinfo {author} {\bibfnamefont {R.~L.}\ \bibnamefont
  {Arnowitt}}, \bibinfo {author} {\bibfnamefont {S.~D.}\ \bibnamefont {Deser}},
  \ and\ \bibinfo {author} {\bibfnamefont {C.~W.}\ \bibnamefont {Misner}},\
  }\href@noop {} {\emph {\bibinfo {title} {The dynamics of general
  relativity}}},\ \bibinfo {type} {Tech. Rep.}\ (\bibinfo {year}
  {1962})\BibitemShut {NoStop}%
\bibitem [{\citenamefont {Schwartz}(2014)}]{schwartz2014quantum}%
  \BibitemOpen
  \bibfield  {author} {\bibinfo {author} {\bibfnamefont {M.~D.}\ \bibnamefont
  {Schwartz}},\ }\href@noop {} {\emph {\bibinfo {title} {Quantum field theory
  and the standard model}}}\ (\bibinfo  {publisher} {Cambridge University
  Press},\ \bibinfo {year} {2014})\BibitemShut {NoStop}%
\bibitem [{\citenamefont {Torres}\ \emph {et~al.}(2018)\citenamefont {Torres},
  \citenamefont {Coutant}, \citenamefont {Dolan},\ and\ \citenamefont
  {Weinfurtner}}]{torres2018waves}%
  \BibitemOpen
  \bibfield  {author} {\bibinfo {author} {\bibfnamefont {T.}~\bibnamefont
  {Torres}}, \bibinfo {author} {\bibfnamefont {A.}~\bibnamefont {Coutant}},
  \bibinfo {author} {\bibfnamefont {S.}~\bibnamefont {Dolan}}, \ and\ \bibinfo
  {author} {\bibfnamefont {S.}~\bibnamefont {Weinfurtner}},\ }\href@noop {}
  {\bibfield  {journal} {\bibinfo  {journal} {Journal of Fluid Mechanics}\
  }\textbf {\bibinfo {volume} {857}},\ \bibinfo {pages} {291} (\bibinfo {year}
  {2018})}\BibitemShut {NoStop}%
\bibitem [{\citenamefont {{Richartz}}\ \emph {et~al.}(2013)\citenamefont
  {{Richartz}}, \citenamefont {{Prain}}, \citenamefont {{Weinfurtner}},\ and\
  \citenamefont {{Liberati}}}]{richartz2013dispersive}%
  \BibitemOpen
  \bibfield  {author} {\bibinfo {author} {\bibfnamefont {M.}~\bibnamefont
  {{Richartz}}}, \bibinfo {author} {\bibfnamefont {A.}~\bibnamefont {{Prain}}},
  \bibinfo {author} {\bibfnamefont {S.}~\bibnamefont {{Weinfurtner}}}, \ and\
  \bibinfo {author} {\bibfnamefont {S.}~\bibnamefont {{Liberati}}},\
  }\href@noop {} {\bibfield  {journal} {\bibinfo  {journal} {Classical and
  Quantum Gravity}\ }\textbf {\bibinfo {volume} {30}},\ \bibinfo {pages}
  {085009} (\bibinfo {year} {2013})}\BibitemShut {NoStop}%
\bibitem [{\citenamefont {Torres}(2020)}]{torres2020estimate}%
  \BibitemOpen
  \bibfield  {author} {\bibinfo {author} {\bibfnamefont {T.}~\bibnamefont
  {Torres}},\ }\href@noop {} {\bibfield  {journal} {\bibinfo  {journal} {arXiv
  preprint arXiv:2003.02230}\ } (\bibinfo {year} {2020})}\BibitemShut {NoStop}%
\bibitem [{\citenamefont {Abramowitz}\ and\ \citenamefont
  {Stegun}(1965)}]{abramowitz1965handbook}%
  \BibitemOpen
  \bibfield  {author} {\bibinfo {author} {\bibfnamefont {M.}~\bibnamefont
  {Abramowitz}}\ and\ \bibinfo {author} {\bibfnamefont {I.~A.}\ \bibnamefont
  {Stegun}},\ }\href@noop {} {\emph {\bibinfo {title} {Handbook of mathematical
  functions: with formulas, graphs, and mathematical tables}}},\ Vol.~\bibinfo
  {volume} {55}\ (\bibinfo  {publisher} {Courier Corporation},\ \bibinfo {year}
  {1965})\BibitemShut {NoStop}%
\bibitem [{\citenamefont {Sotiriou}\ \emph {et~al.}(2011)\citenamefont
  {Sotiriou}, \citenamefont {Visser},\ and\ \citenamefont
  {Weinfurtner}}]{sotiriou2011lower}%
  \BibitemOpen
  \bibfield  {author} {\bibinfo {author} {\bibfnamefont {T.~P.}\ \bibnamefont
  {Sotiriou}}, \bibinfo {author} {\bibfnamefont {M.}~\bibnamefont {Visser}}, \
  and\ \bibinfo {author} {\bibfnamefont {S.}~\bibnamefont {Weinfurtner}},\
  }\href {\doibase 10.1103/PhysRevD.83.124021} {\bibfield  {journal} {\bibinfo
  {journal} {Physical Review D}\ }\textbf {\bibinfo {volume} {83}},\ \bibinfo
  {pages} {124021} (\bibinfo {year} {2011})}\BibitemShut {NoStop}%
\bibitem [{\citenamefont {Barausse}\ \emph {et~al.}(2011)\citenamefont
  {Barausse}, \citenamefont {Jacobson},\ and\ \citenamefont
  {Sotiriou}}]{barausse2011black}%
  \BibitemOpen
  \bibfield  {author} {\bibinfo {author} {\bibfnamefont {E.}~\bibnamefont
  {Barausse}}, \bibinfo {author} {\bibfnamefont {T.}~\bibnamefont {Jacobson}},
  \ and\ \bibinfo {author} {\bibfnamefont {T.~P.}\ \bibnamefont {Sotiriou}},\
  }\href {\doibase 10.1103/PhysRevD.83.124043} {\bibfield  {journal} {\bibinfo
  {journal} {Physical Review D}\ }\textbf {\bibinfo {volume} {83}},\ \bibinfo
  {pages} {124043} (\bibinfo {year} {2011})}\BibitemShut {NoStop}%
\bibitem [{\citenamefont {Barausse}\ and\ \citenamefont
  {Sotiriou}(2013)}]{barausse2013black}%
  \BibitemOpen
  \bibfield  {author} {\bibinfo {author} {\bibfnamefont {E.}~\bibnamefont
  {Barausse}}\ and\ \bibinfo {author} {\bibfnamefont {T.~P.}\ \bibnamefont
  {Sotiriou}},\ }\href@noop {} {\bibfield  {journal} {\bibinfo  {journal}
  {Classical and Quantum Gravity}\ }\textbf {\bibinfo {volume} {30}},\ \bibinfo
  {pages} {244010} (\bibinfo {year} {2013})}\BibitemShut {NoStop}%
\bibitem [{\citenamefont {Patrick}\ \emph {et~al.}(2018)\citenamefont
  {Patrick}, \citenamefont {Coutant}, \citenamefont {Richartz},\ and\
  \citenamefont {Weinfurtner}}]{patrick2018quasibound}%
  \BibitemOpen
  \bibfield  {author} {\bibinfo {author} {\bibfnamefont {S.}~\bibnamefont
  {Patrick}}, \bibinfo {author} {\bibfnamefont {A.}~\bibnamefont {Coutant}},
  \bibinfo {author} {\bibfnamefont {M.}~\bibnamefont {Richartz}}, \ and\
  \bibinfo {author} {\bibfnamefont {S.}~\bibnamefont {Weinfurtner}},\ }\href
  {\doibase 10.1103/PhysRevLett.121.061101} {\bibfield  {journal} {\bibinfo
  {journal} {Physical Review Letters}\ }\textbf {\bibinfo {volume} {121}},\
  \bibinfo {pages} {061101} (\bibinfo {year} {2018})}\BibitemShut {NoStop}%
\bibitem [{\citenamefont {Churilov}\ and\ \citenamefont
  {Stepanyants}(2019)}]{churilov2018scattering}%
  \BibitemOpen
  \bibfield  {author} {\bibinfo {author} {\bibfnamefont {S.}~\bibnamefont
  {Churilov}}\ and\ \bibinfo {author} {\bibfnamefont {Y.}~\bibnamefont
  {Stepanyants}},\ }\href@noop {} {\bibfield  {journal} {\bibinfo  {journal}
  {Physical Review Fluids}\ }\textbf {\bibinfo {volume} {4}},\ \bibinfo {pages}
  {034704} (\bibinfo {year} {2019})}\BibitemShut {NoStop}%
\bibitem [{\citenamefont {Lucassen}(1968)}]{lucassen1968longitudinal1}%
  \BibitemOpen
  \bibfield  {author} {\bibinfo {author} {\bibfnamefont {J.}~\bibnamefont
  {Lucassen}},\ }\href@noop {} {\bibfield  {journal} {\bibinfo  {journal}
  {Transactions of the Faraday Society}\ }\textbf {\bibinfo {volume} {64}},\
  \bibinfo {pages} {2221} (\bibinfo {year} {1968})}\BibitemShut {NoStop}%
\bibitem [{\citenamefont {LeBlond}\ and\ \citenamefont
  {Mainardi}(1987)}]{leblond1987viscous}%
  \BibitemOpen
  \bibfield  {author} {\bibinfo {author} {\bibfnamefont {P.~H.}\ \bibnamefont
  {LeBlond}}\ and\ \bibinfo {author} {\bibfnamefont {F.}~\bibnamefont
  {Mainardi}},\ }\href@noop {} {\bibfield  {journal} {\bibinfo  {journal} {Acta
  Mechanica}\ }\textbf {\bibinfo {volume} {68}},\ \bibinfo {pages} {203}
  (\bibinfo {year} {1987})}\BibitemShut {NoStop}%
\bibitem [{\citenamefont {Alpers}\ and\ \citenamefont
  {H{\"u}hnerfuss}(1989)}]{alpers1989damping}%
  \BibitemOpen
  \bibfield  {author} {\bibinfo {author} {\bibfnamefont {W.}~\bibnamefont
  {Alpers}}\ and\ \bibinfo {author} {\bibfnamefont {H.}~\bibnamefont
  {H{\"u}hnerfuss}},\ }\href@noop {} {\bibfield  {journal} {\bibinfo  {journal}
  {Journal of Geophysical Research: Oceans}\ }\textbf {\bibinfo {volume}
  {94}},\ \bibinfo {pages} {6251} (\bibinfo {year} {1989})}\BibitemShut
  {NoStop}%
\bibitem [{\citenamefont {Przadka}\ \emph {et~al.}(2012)\citenamefont
  {Przadka}, \citenamefont {Cabane}, \citenamefont {Pagneux}, \citenamefont
  {Maurel},\ and\ \citenamefont {Petitjeans}}]{przadka2012fourier}%
  \BibitemOpen
  \bibfield  {author} {\bibinfo {author} {\bibfnamefont {A.}~\bibnamefont
  {Przadka}}, \bibinfo {author} {\bibfnamefont {B.}~\bibnamefont {Cabane}},
  \bibinfo {author} {\bibfnamefont {V.}~\bibnamefont {Pagneux}}, \bibinfo
  {author} {\bibfnamefont {A.}~\bibnamefont {Maurel}}, \ and\ \bibinfo {author}
  {\bibfnamefont {P.}~\bibnamefont {Petitjeans}},\ }\href@noop {} {\bibfield
  {journal} {\bibinfo  {journal} {Experiments in Fluids}\ }\textbf {\bibinfo
  {volume} {52}},\ \bibinfo {pages} {519} (\bibinfo {year} {2012})}\BibitemShut
  {NoStop}%
\end{thebibliography}%

\appendix
\numberwithin{equation}{section}
\section{Numerical solution in shallow water} \label{app:num}

To obtain an exact spectrum from the reflection coefficient in shallow water, we perform a numerical simulation following \cite{churilov2018scattering}.
To lighten the notation, we define the dimensionless quantities,
\begin{equation} \label{adim}
\sigma = \frac{\omega D}{c^2}, \qquad B = \frac{C}{D}, \qquad x = \frac{rc}{D}, \qquad \tau = \frac{tc^2}{D}.
\end{equation}
The wave equation \eqref{wave_equation} with the shallow water dispersion function \eqref{shallowF} may then be written,
\begin{equation} \label{waveeqnFx}
\begin{split}
x^2(x^2-1)f'' + & \ \left[1+x^2-2i\left(\sigma x^2-mB\right)\right]xf' \\
& + \left[(\sigma x^2-mB)^2-2imB-m^2x^2\right]f = 0,
\end{split}
\end{equation}
where $'=\partial_x$ and we have written the perturbation for a particular \mbox{$(m,\sigma)$} mode using the ansatz \mbox{$\phi = \mathrm{Re}[f(x)\exp(im\theta-i\sigma\tau)]$}.
Noticing the appearance of powers of $x^2$ in this equation, we define a new variable $y=x^2$ which leads to,
\begin{equation} \label{waveeqnFy}
\begin{split}
y^2(y-1)\partial_y^2f+ & \ \left[y-i(\sigma y-mB)\right]y\partial_yf \\
& + \frac{1}{4}\left[(\sigma y-mB)^2-2imB-m^2y\right]f = 0.
\end{split}
\end{equation}
Both Eqs.~\eqref{waveeqnFx} and \eqref{waveeqnFy} are second order ordinary differential equations with a regular singular point at $x=1$ and $y=1$ respectively \cite{churilov2018scattering}.
A numerical solution requires initial conditions which are provided by the Frobenius expansion of $f$ about the regular singular point.
This is most easily obtained from Eq.~\eqref{waveeqnFy}, since the polynomials in front of $f$ and it's derivatives are of lower order.
We first write $f$ as,
\begin{equation}
f(y) = (y-1)^p\sum_{n=0}^\infty a_n(y-1)^n,
\end{equation}
and substitute into Eq.~\eqref{waveeqnFy}.
Demanding that the equation is satisfied for the lowest order in $(y-1)$, the index is given by,
\begin{equation}
p = \begin{cases}
0 \\
i(\sigma-mB)
\end{cases},
\end{equation}
The different values of $p$ correspond to the two linearly independent solutions.
The solution with $p=i(\sigma-mB)$ is an out-going mode which diverges on the horizon.
To see this, one can write the overall factor preceding the power series $(y-1)^p=\exp(i(\sigma-mB)\log(y-1))$ which has the form of a plane wave whose wave number diverges as $y\to1$.
Indeed, this solution corresponds to the mode we discarded in Eq.~\eqref{shallow_asymp} when we imposed a purely in-going boundary condition on the horizon.
Hence, we discard it for the same reason here.
The next lowest order in $(y-1)$ gives an expression for $a_1$ in terms of $a_0$, but since Eq.~\eqref{waveeqnFy} is linear in $f$ we may set $a_0=1$.
Hence the first two terms in the expansion are,
\begin{equation} \label{frob}
f(y) = 1 - \frac{(\sigma-mB)^2-2imB-m^2}{4[1-i(\sigma-mB)]}(y-1) + \mathcal{O}\left((y-1)^2\right).
\end{equation}
From Eq.~\eqref{frob}, we convert back to the $x$ variable using \mbox{$f'(x=1)=2\partial_yf|_{y=1}$}, then compute initial conditions \mbox{$f(1+\epsilon)$} to \mbox{$\mathcal{O}(\epsilon^2)$} and \mbox{$f'(1+\epsilon)$} to $\mathcal{O}(\epsilon)$.
Better accuracy can be obtained by taking higher order terms in the expansion. 
However, we found this was not necessary since we were able to find consistent solutions for $\epsilon=10^{-4}$ and $\epsilon=10^{-6}$ (an error due to poor initial conditions would decrease with $\epsilon$).
These initial conditions are used to solve Eq.~\eqref{waveeqnFx} numerically over the range $x\in[1+\epsilon,x_n]$ with $x_n = 20(2\pi/\sigma)$, i.e. about $20$ (flat space) wavelengths away from the centre.
We used Matlab's inbuilt function \emph{ode45} (which is based on a forth order Runge-Kutta algorithm) to evolve from the starting point into the asymptotic region.

To extract the amplitudes in this region, we use the asymptotic solution to Eq.~\eqref{waveeqnFx}:
\begin{equation} \label{fSolInfty}
f(x\to\infty) = x^{i\sigma-1/2}\left(A_\infty^-e^{-i\sigma x} + A_\infty^+e^{i\sigma x} \right).
\end{equation}
Using Eq.~\eqref{fSolInfty} and it's derivative, we may solve for $A_\infty^\pm$ in terms of our numerical solution at $x=x_n$, i.e. $(f_n,f'_n)$.
This gives,
\begin{equation}
\frac{A_\infty^+}{A_\infty^-} = -e^{-2i\sigma x_n}\frac{2x_nf'_n+f_n(1-2i\sigma(1-x_n))}{2x_nf'_n+f_n(1-2i\sigma(1+x_n))},
\end{equation}
which is just the reflection coefficient $\mathcal{R}$ in \eqref{shallow_coefs}.
The frequency dependence of $\mathcal{R}$ for different $m$ is shown in Fig.~\ref{fig:Shallow} and later in Figs.~\ref{fig:Compare} and \ref{fig:LowRotation} where it is compared to the deep water result.

\section{Reflection coefficient in deep water} \label{app:scattering}

In this appendix, we detail the derivation of the reflection coefficients in \eqref{refl_type1} to \eqref{refl_type6} from the scattering matrix $\mathcal{M}$ in \eqref{scatterMat}.
The scattering matrix is assisted by looking at the Feynman diagrams in Fig.~\ref{fig:Feynmann_super} 
and proceeds as follows.
Starting from the mode amplitudes at $r=R$, each mode is evolved inward using the shift factor $\SH_{ab}$ defined in \eqref{WKBoperator}.
Each time two modes interact and then depart, there will be a multiplication by $\mathcal{N}_{ab}$ defined in \eqref{LocalScatter2}.
When two modes are interacting all the way down to $r=0$, there will be a multiplication by $\widetilde{T}$ defined in \eqref{2tp_cf}.
The resulting matrix equation can then be solved for the scattering coefficients.
The calculation is analogous to that in Section~\ref{sec:shal_refl} 
for shallow water, albeit with 2 extra modes and the possibility of more interactions.

As an example, we sketch the computation for type V scattering.
The $4\times 4$ scattering matrix is,
\begin{widetext}
\begin{equation} \label{typeVmatrix}
\mathcal{M} = \begin{pmatrix}
1 &   &  \\
  & 1 & \\
  &   & \widetilde{T}
\end{pmatrix} \begin{pmatrix}
\SH^\mathrm{u}_{12} & & & \\
 & \SH^+_{12} & & \\
 & & \SH^-_{12} & \\
 & & & \SH^\mathrm{d}_{12} 
\end{pmatrix} \begin{pmatrix}
\SH^\mathrm{u}_{12} & & \\
 & \SH^\downarrow_{23}\mathcal{N}_{23} & \\
 & & \SH^\mathrm{d}_{23}
\end{pmatrix} \begin{pmatrix}
\SH^\mathrm{u}_{3R} & & & \\
 & \SH^+_{3R} & & \\
 & & \SH^-_{3R} & \\
 & & & \SH^\mathrm{d}_{3R} 
\end{pmatrix}.
\end{equation}
\end{widetext}
The mode amplitudes are then related by,
\begin{equation}
\begin{pmatrix}
A_1^\mathrm{u} \\ A_1^+ \\ A_1^\downarrow \\ 0
\end{pmatrix} = \mathcal{M} \begin{pmatrix}
A_R^\mathrm{u} \\ A_R^+ \\ A_R^- \\ A_R^\mathrm{d}
\end{pmatrix}.
\end{equation}
where we have set the amplitude of the mode which grows toward to centre to zero, since a divergence at $r=0$ is non-physical.
Now, we are looking for an equation which relates the mode amplitudes at infinity from which we can apply \eqref{R_coeff} to compute the reflection coefficient.
Clearly, this can be obtained by evaluating the bottom row of $\mathcal{M}$ and finding the dot product with the amplitude vector.
However, the matrix multiplications involved are rather lengthy and tedious.
A simpler method involves evaluating the amplitudes at the intermediate locations as follows.

The matrix on the far left \eqref{typeVmatrix} gives a relation between the amplitudes of the d and $-$ modes are $r_1$,
\begin{equation} \label{amp1}
A_1^\mathrm{d} = iA_1^-.
\end{equation}
Next, one can see from \eqref{typeVmatrix} that the d mode evolves adiabatically between $r_1$ and $R$, whereas the $+$ and $-$ evolve adiabatically between $r_3$ and $R$:
\begin{equation} \label{amp2}
\begin{split}
A_1^\mathrm{d} = & \ \SH^\mathrm{d}_{1R}A_R^\mathrm{d}, \\
A_3^+ = & \ \SH^+_{3R}A_R^+ \\
A_3^- = & \ \SH^-_{3R}A_R^- \\
\end{split}
\end{equation}
The relation between $A^\pm_3$ and $A^-_2$ is provided by evaluating the bottom row of $\mathcal{N}_{23}$,
\begin{equation} \label{amp3}
A_2^- = \SH^\downarrow_{23}\left[-i\left(1-\tfrac{1}{4}f_{23}^2\right)A_3^+ + \left(1+\tfrac{1}{4}f_{23}^2\right)A_3^-\right],
\end{equation}
with $f_{12}$ given in \eqref{LocalScatter2}, and finally $A^-_1$ is adiabatically related to $A^-_2$,
\begin{equation} \label{amp4}
A^-_1 = \SH^-_{12}A^-_2.
\end{equation}
The equations in \eqref{amp1}-\eqref{amp4} constitute 6 relations between 8 unknown coefficients.
These can be combined to give a single relation between the amplitudes of the d,$-$ and $+$ modes at $r=R$.
Using the definition of the coefficients $\mathcal{R}$ and $\mathcal{I}^\mathrm{d}$ in \eqref{R_coeff} and \eqref{I_coeff}, this yields the expression in \eqref{refl_type5}.
The $e^{i\varphi^j}$ terms result from the combination of the different WKB phases appearing in the $\SH^j_{ab}$.
The calculation proceeds analogously for the other 5 scattering types and results in the expressions for $\mathcal{R}$ quoted in the main text.

\section{Dissipation} \label{app:diss}

One of the major challenges in predicting spectrum of the reflection coefficient is in modelling the effects of dissipation.
Dissipation of the waves could either be due to the viscosity of the fluid, or dissipation into other types of waves.
Here, we show how these effects can be included in the dispersion relation for waves propagating the $x$ direction through a static fluid.

The derivation of the dispersion relation closely follows that of Lucassen \cite{lucassen1968longitudinal1}.
In two dimensions, the linearised Navier-Stokes equations for velocity perturbations $\mathbf{v}$ can be brought to the form \cite{leblond1987viscous},
\begin{equation} \label{LinNS}
\nabla^2\phi = 0, \qquad \partial_t\psi = \nu\nabla^2\psi,
\end{equation}
where the scalar potential $\phi$ and stream function $\psi$ have been introduced via,
\begin{equation}
\begin{split}
v_x = & \ \partial_x\phi - \partial_z\psi, \\
v_z = & \ \partial_z\phi + \partial_x\psi.
\end{split}
\end{equation}
Using the gauge freedom, $\phi$ can be chosen to satisfy Bernoulli's equation,
\begin{equation}
\partial_t\phi + \tfrac{1}{\rho}p + gz = 0,
\end{equation}
where $p$ is the pressure and $\rho$ is the fluid's density. The boundary conditions at  $z=-h$ are the no-slip and no-penetration conditions,
\begin{equation} \label{BCfloor}
\mathbf{v}(z=-h) = 0.
\end{equation}
At the water's surface $z=0$, the linearised boundary conditions for the normal and tangential stresses are \cite{alpers1989damping},
\begin{subequations}
\begin{align}
\partial_t\phi + g\eta - 2\nu\partial_zv_z - \tfrac{1}{\rho}\gamma\partial_x^2\eta = 0 & , \label{BCtop1} \\
\tfrac{1}{\rho}\partial_x\gamma = \nu\left(\partial_zv_x+\partial_xv_z\right)|_{z=0} & , \label{BCtop2} \\
v_z(z=0) = \partial_t\eta & , \label{BCtop3}
\end{align}
\end{subequations}
where $\eta$ is a free surface fluctuation and $\gamma$ is the surface tension.
These boundary conditions include the possibility of spatial variations in surface tension through the $\partial_x\gamma$ term.
This can be related to the horizontal fluid displacement $\xi$, which satisfies $v_x=\partial_t\xi$ at linear order, via \cite{alpers1989damping},
\begin{equation}
\partial_x\gamma = E\partial_x^2\xi,
\end{equation}
where $E$ is the surface dilational modulus.
Surface tension gradients are responsible for the existence of a type of longitudinal waves called Marangoni waves \cite{przadka2012fourier}.

In the deep water regime, one may set $h\to-\infty$ and the solutions to \eqref{LinNS} of frequency $\omega$ satisfying \eqref{BCfloor} are,
\begin{align}
\phi = & \ A e^{kz}e^{ikx-i\omega t}, \label{sol1} \\
\psi = & \ B e^{lz}e^{ikx-i\omega t}, \label{sol2}
\end{align}
where $l$ is determined by,
\begin{equation}
l^2 = k^2 - i\omega/\nu.
\end{equation}
In the limit of vanishing viscosity, the solutions are completely characterised by the potential function.
Thus when viscosity is small, one expects $A\gg B$.
Furthermore, the stream function is effectively confined to the free surface since $l\sim \nu^{-\frac{1}{2}}$ is large.
The dispersion relation for a free surface fluctuation $\eta$ is obtained by substituting solutions \eqref{sol1} and \eqref{sol2} into the boundary conditions at $z=0$.
The first boundary condition \eqref{BCtop1} becomes,
\begin{equation}
\omega^2 - gk - \sigma k^3 + 2i\nu\omega k^2 = \frac{B}{A}(2\nu\omega l k - igk - i\sigma k^3).
\end{equation}
Notice that this reduces to the usual dispersion relation for capillary-gravity waves when $\nu=0$ and $B=0$.
When viscosity is included at the free surface, but vorticity is neglected, the left hand side vanishes which gives a known modification of the dispersion relation, e.g. \cite{torres2018waves}.
The second boundary condition \eqref{BCtop2} can be recast as an expression for the ratio $B/A$,
\begin{equation}
\frac{B}{A} = \frac{2i\nu\omega k^2 + Ek^3}{iElk^2-\nu\omega(l^2+k^2)} \overset{\nu\to 0}{\sim} \sqrt{\frac{i\nu}{\omega}}k + \mathcal{O}(\nu),
\end{equation}
which shows that when viscosity is weak, the contribution of vorticity to the fluid motion is indeed small.
Inserting the approximate form of $B/A$ into the first boundary condition yields the following dispersion relation,
\begin{equation} \label{dispersion_damp}
\omega^2 = \left(1-\sqrt{\frac{i\nu}{\omega}}k\right)(gk+\sigma k^3) + \mathcal{O}(\nu),
\end{equation}
where $k$ is complex.
Let this be written $k=\kappa+i\beta$ where $\beta$ is the damping coefficient satisfying $|\beta|\ll|\kappa|$.
Neglecting surface tension, a leading order approximation gives,
\begin{equation} \label{apprx1}
\omega^2 = g\kappa, \qquad \beta = \sqrt{\frac{\nu}{2\omega}}g\kappa^2.
\end{equation}
Note that the surface dilational modulus $E$ has dropped out of the expression and thus, the present approximation does not include the coupling of surface waves to Marangoni waves.
This can easily be included by adding the next to leading order term in $\nu$.

To check the validity of these approximations, consider the experimental data in \cite{torres2017rotational}.
In their Fig. 3, a plane wave propagating in standing water has a reflection coefficient of around $\mathcal{R}\sim 0.76$ across all azimuthal components.
Since radial waves are completely reflected in standing water, the reflection coefficient is influenced only by the amount of damping that occurs over the distance travelled, which is roughly $x_\mathrm{tot}\sim 0.5~\mathrm{m}$.
The approximation in \eqref{apprx1} gives $\mathcal{R}=e^{-\beta x_\mathrm{tot}}\sim 0.8$ which is in good agreement with experiment.

The result in \eqref{dispersion_damp}, which is valid for a static fluid, indicates that dissipation affects short wavelengths more strongly than longer ones.
Given that $p^\mathrm{u}$ continues to grow at large $r$, one would therefore expect any incident u modes to be completely damped out over a relatively short distance, which justifies their neglect in the previous section.
Another potential use of the analysis of this section could be to account for the effects of dissipation on the reflection coefficient in inhomogeneous media.
To do this properly, however, would require a derivation of \eqref{dispersion_damp} for non-zero flow fields, which is beyond the scope of this paper.
Assuming weak viscosity, the effects of dissipation could then be included in the WKB amplitude, as explained in \cite{torres2018waves}.
This would be a worthwhile exploration for future research.

\end{document}